\documentclass[fleqn,usenatbib]{mnras}

\usepackage[T1]{fontenc}

\DeclareRobustCommand{\VAN}[3]{#2}
\let\VANthebibliography\thebibliography
\def\thebibliography{\DeclareRobustCommand{\VAN}[3]{##3}\VANthebibliography}

\usepackage{graphicx}   
\usepackage{amsmath}    
\usepackage{amssymb}    
\usepackage{gensymb}
\usepackage{newtxtext,newtxmath}

\title[ORB6 binaries in Gaia EDR3]{Visual binary stars with known orbits in Gaia EDR3}

\author[Chulkov \& Malkov]{
Dmitry Chulkov,$^{1}$\thanks{chulkovd@gmail.com}
Oleg Malkov$^{1}$
\\
$^{1}$INASAN, 48 Pyatnitskaya St., Moscow 119017, Russia\\}

\date{Accepted XXX. Received YYY; in original form ZZZ}

\pubyear{2015}

\begin{document}
\label{firstpage}
\pagerange{\pageref{firstpage}--\pageref{lastpage}}
\maketitle

\begin{abstract}

3350 objects from the Sixth catalog of orbits of visual binary stars
(ORB6) are investigated to validate \textit{Gaia} EDR3 parallaxes
and provide mass estimates for the systems. We show that 2/3 of
binaries with 0.2 -- 0.5 arcsec separation are left without a
parallax solution in EDR3. A special attention is paid to 521
pairs with parallax known separately for both components. We find
16 entries that are deemed
to be chance alignments of unrelated stars. At once we show examples of high-confidence binary systems with significant
differences in the reported parallaxes of their components. Next we conclude that the reported Gaia EDR3 parallax errors are
underestimated, at least by a factor of 3 for sources with large RUWE. Parallaxes are needed to estimate stellar masses. Since nearly 30\% of ORB6 entries lack 5 or 6-parameter solution in
EDR3, we attempt to enrich the astrometric data. Distant
companions of ORB6 entries are revealed in EDR3 by analysis of stellar proper motions and \textit{Hipparcos} parallaxes. Notably, in 28 cases intrinsic EDR3 parallaxes of the binary components appear to be less reliable than the parallax of the outer companions.
\textit{Gaia} DR2, TGAS and \textit{Hipparcos} parallaxes are used
when EDR3 data is unavailable. Synthetic mass-luminosity relation
in the $G$ band for main sequence stars is obtained to provide mass estimates along with dynamical masses calculated via Kepler's
Third Law.
\end{abstract}


\section{Introduction}

A pair of stars which appears close in the sky is known as a
double star. \citet{1803RSPT...93..339H}, using observations of
Castor ($\alpha$ Gem) made in 1759 -- 1803, concluded that
continual change of the position angle could not be explained by
the stellar proper motions. Instead, Castor together with a small
companion move round their common centre of gravity. Nowadays we
attribute this pair to the class of visual binary stars. Long
observational sets with measurements of three parameters: time of
observation, position angle, and separation between the components
are needed to calculate the apparent and true orbit. This problem
was analytically solved by \citet{Savary1827}. The orbital
elements that can be derived include period $P$ and semi-major
axis $a^{\prime\prime}$, deduced in angular units. Knowledge of
parallax $\varpi$ is required to convert $a^{\prime\prime}$ to
linear measure, $a \sim \frac{a^{\prime\prime}}{\varpi}$. Then,
Kepler's Third Law $M_{d} \sim \frac{a^3}{P^2} \sim
\frac{a^{\prime\prime3}}{\varpi^3 P^2}$ allows one to calculate
the total mass of the resolved binary. Due to large orbital
periods and scarce parallax data, applicability of this method has
been rather limited: while $\citet{}$ \citet{1837sdmm.book.....S}
catalogue contains 2714 double stars, \citet{1918bist.book.....A}
lists 112 visual binaries with known orbits, with masses
estimated for just 14 of them. The \textit{Hipparcos} mission
\citep{1997A&A...323L..49P} provided reliable parallaxes for
hundreds of visual binaries; still, the double-lined eclipsing
binaries remain the prime source of precise stellar masses
\citep{1980ARA&A..18..115P,
2010A&ARv..18...67T,2021A&ARv..29....4S}.

\textit{Gaia} space mission
\citep{2016A&A...595A...1G} is expected to make a breakthrough and
has already brought nearly 0.8 million non-single stars including
more than 165 thousand astrometric orbital solutions
\citep{2022arXiv220605726H} in its third data release \citep{2022arXiv220800211G}. Still, it
represents a small fraction of binary population; particularly,
99\% of the processed systems are in  the 0.28 -- 1500 day period
range \citep{2022arXiv220605595G}. Largely, new non-single-star
data do not overlap the long-existing observational results for
visual binaries treated in this paper. \textit{Gaia} DR3
essentially complements the EDR3 catalogue
\citep{2021A&A...649A...1G}, and the content of the main table,
including parallaxes, remains unaltered. It means that the
presented results are completely relevant despite being based on
\textit{Gaia} EDR3. The information about known orbits of visual
binaries is collected in the Sixth Catalog of Orbits of Visual
Binary Stars, ORB6 \citep{2001AJ....122.3472H}, maintained by the
US Naval observatory along with the Washington Double Star
catalogue \citep{2001AJ....122.3466M} which is a principal
catalogue for all visual binaries. ORB6 does not provide stellar
parallaxes needed for mass calculation.
\citet{2012A&A...546A..69M} combined the available
\textit{Hipparcos} and ORB6 data to estimate masses. Recent
\textit{Gaia} data releases significantly improved our knowledge
of stellar distances. Therefore, we attempt to supply ORB6 orbits
with new available parallaxes in the present paper.

In the next Section, we briefly describe the ORB6 catalogue and
our efforts to find \textit{Gaia} counterparts for its objects. In
Section \ref{sec:2plx}, resolved double stars are investigated to
reveal optical pairs among them and validate EDR3 parallaxes. In
Section \ref{sec:third}, we search for outer components of ORB6
binaries to expand available parallax data. Section
\ref{sec:outer} shortly describes data retrieved from other
catalogues. In Section \ref{sec:masses}, two methods for
estimating stellar masses are presented. The obtained results are
discussed in Section \ref{sec:comparison}. The conclusions in
Section \ref{sec:conclusions} are followed by Appendix
\ref{sec:appendix} containing the complete tabulated data.

\section{ORB6 binaries and identification in \textit{Gaia} EDR3}
\label{sec:edr3} As of April 2022, ORB6 provides coordinates and
orbital elements, including orbital period $P$ and semi-major axis
in the angular measure $a^{\prime\prime}$, for 3460 entries.
Apparent magnitudes useful for identification in outer data sets
are available, however, these data generally are less consistent.
Orbits are graded on a scale of 1--5 according to their quality,
with grade 1 representing definitive orbits while grade 5 is
reserved for the least reliable (undeterminate) solutions;
additionally, 7 is assigned for systems with incomplete orbital
elements and 8 is used for interferometric binaries. Astrometric
binaries are marked by grade 9. Unfortunately, for the latter
class, the perturbation amplitude \citep{2017AJ....153..258B} is
often published instead of $a^{\prime\prime}$ which should not be
applied in Kepler's Third Law, and its accidental use will cause
spurious mass estimates. Dubious entries occasionally appear in
ORB6 among other grades as well. We choose to keep all suspicious
entries in the sample, they are manifested by unrealistic
dynamical mass in Table \ref{tab:master}.

ORB6 binaries are rather diverse, covering at least 3 orders of
magnitude for $P$ and $a^{\prime\prime}$. Unfortunately, more than
40\% of entries lack error estimates. For such systems, we
estimate relative uncertainty of the orbit size and period
according to the 75\% quartile for its corresponding orbit grade.
Several alternative solutions for one pair and multiple systems
appear in the catalogue, notably 42 orbits are related to the Sgr
$A^*$ cluster at the Galactic Center, WDS 17457-2900. The overall
statistics describing ORB6 binaries as a function of their quality
grade is summed up in Table \ref{tab:grades}. Along with orbital
elements, predictions of angular separation $\rho_{\rm eph}$ and
position angle for a given date are available; the authors thank
Rachel Matson for providing the ephemerides at the \textit{Gaia}
EDR3 epoch.

\begin{table*}
    \centering
\caption{General properties of ORB6 binary stars as a function of
the orbit's quality grade, see Section \ref{sec:edr3}. Line 1
indicates the total number of orbital solutions of the given
grade. Occasionally, several solutions for the same pair are
present; for subsequent statistics, only one entry per pair is
chosen. The number of the unique pairs is shown in line 2. Pairs
with successful cross-matching in \textit{Gaia} EDR3 are summed up
in line 3. Line 4 counts systems with EDR3 parallax for at least
one component. Line 5 counts systems with primary and secondary
components both appearing as separate sources in EDR3. Line 6
shows the number of systems with parallaxes available for both
components. Line 7 shows the median orbital period according to
ORB6. Line 8 is the 75\% quartile of the relative period error,
this value is used when errors are not provided for a given
binary. Lines 9--10 show the median semimajor axis provided in
ORB6 and its 75\%-quartile relative error. Lines 11--13 list the
median apparent $G$ EDR3 magnitude, parallax, and relative
parallax error of the primary components.}
    \begin{tabular}{ccccccccccc}
       ü& Grade & 1 &2 & 3 & 4 & 5& 7 &  8 & 9 & $\Sigma$ \\
                \hline
1&Orbits &90&384&717&997&669&41&19&543&3460\\
2&Unique pairs&90&382&713&973&628&40&18&506&3350\\
3&Identified pairs&88&370&706&956&612&39&18&490&3279\\
4&$\varpi_1$ or $\varpi_2$ &40&188&390&662&506&35&17&425&2263\\
5&Resolved pairs&10&34&91&295&321&--&--&2&753\\
6&$\varpi_1$ and $\varpi_2$ &4&16&50&200&250&--&--&1&521 \\
7&Average $P$, years&11.0&26.5&67.6&217&450&18.0&0.05&2.49&80 \\
8&3/4 quartile error, \% &0.1&0.9&5.0&40&49&26&< 0.1&8.9&15 \\
9&Average $a$, arcsec&0.17&0.19&0.24&0.50&1.18&0.12&0.005&0.01&0.29 \\
10&3/4 quartile error, \% & 0.9&2.1&5.6&25&52&4.6&3.4&25&19\\
11&Average primary $G$ magnitude &5.5&7.0&7.9&8.0&8.3&8.9&5.2&6.7&7.7\\
12&Average parallax $\varpi_1$, mas&21.7&19.4&12.1&12.1&14.0&13.5&27.5&16.3&14.4\\
13&Average error $\sigma_1/\varpi_1$,  \%&1.6&1.6&1.9&1.1&0.5&0.2&0.4&2.0&1.3\\
\hline
    \end{tabular}
    \label{tab:grades}
\end{table*}

\textit{Gaia} early data release 3 (EDR3) includes 1.8 billion
sources \citep{2021A&A...649A...1G} with coordinates for the epoch
J2016.0. The larger part is constructed from either 5- or
6-parameter astrometric solutions which provide information on
stellar proper motions and parallaxes. The discretion between
these solution types depends on the chromaticity correction method
used in the processing algorithm \citep{2021A&A...649A...2L}.
Throughout this paper, we refer to both types as full solutions.
In EDR3, all sources are assumed to move linearly with uniform
velocity relative to the solar system barycentre, which is
appropriate for single stars. Despite the absence of dedicated
multiple-star solutions, \textit{Gaia} EDR3 data are widely used
for the study of binaries \citep{2021MNRAS.506.2269E}. Indeed,
visual binaries are used for validation of EDR3 data
\citep{2021A&A...649A...4L, 2021A&A...649A...5F}. 2-parameter
solutions which lack parallax and proper motion values are
published for the sources whose full solutions do not converge
well. The quality of the full astrometric solution can be assessed
with the goodness-of-fit parameter RUWE (renormalised unit weight
error) which we denote $\chi$. $\chi \simeq 1.0$ for a
well-behaved sources, and $\chi$ is not available for 2-parameter
solutions. Unresolved binary stars are known to be manifested by a
large RUWE value \citep{2020MNRAS.496.1922B,
2021ApJ...907L..33S}. When several sources within 0.18 arcsec are
found by the processing algorithm, the best one is retained in
EDR3 and marked as duplicated. Along with astrometric data, we use
$G$-band photometry which is available for nearly every source
\citep{2021A&A...649A...3R}.

More than 90\% of ORB6 binaries have apparent magnitude
mag$_1<10.5$ in the V band; therefore, relatively bright ORB6
stars normally stand out among the field of the \textit{Gaia}
sources. Identification of \textit{Gaia} EDR3 counterparts is
based on the epoch J2000.0 coordinates provided in ORB6. The
initial search is done in 1 arcsec radius with the help of TOPCAT
service \citep{2005ASPC..347...29T}. For the yet unmatched
binaries, the search radius is gradually increased. The majority
of the sources are found within 0.5 arcsec from the reference
position. However, for 52 objects, this distance
exceeds 10 arcsec, most of these binaries possess high proper
motions. Stellar magnitudes and WDS data on proper motions are
used for confirmation of correct cross-matching when needed. ORB6
binaries often constitute a part of a multiple system. If more
than one \textit{Gaia} source is associated with the binary, the
position angle and angular separation $\rho$ is calculated.
Cross-identification is carefully evaluated to prevent matching of
the wrong component in the multiple system or a chance alignment
star.

\begin{figure}
    \centering
    \includegraphics[width=\columnwidth]{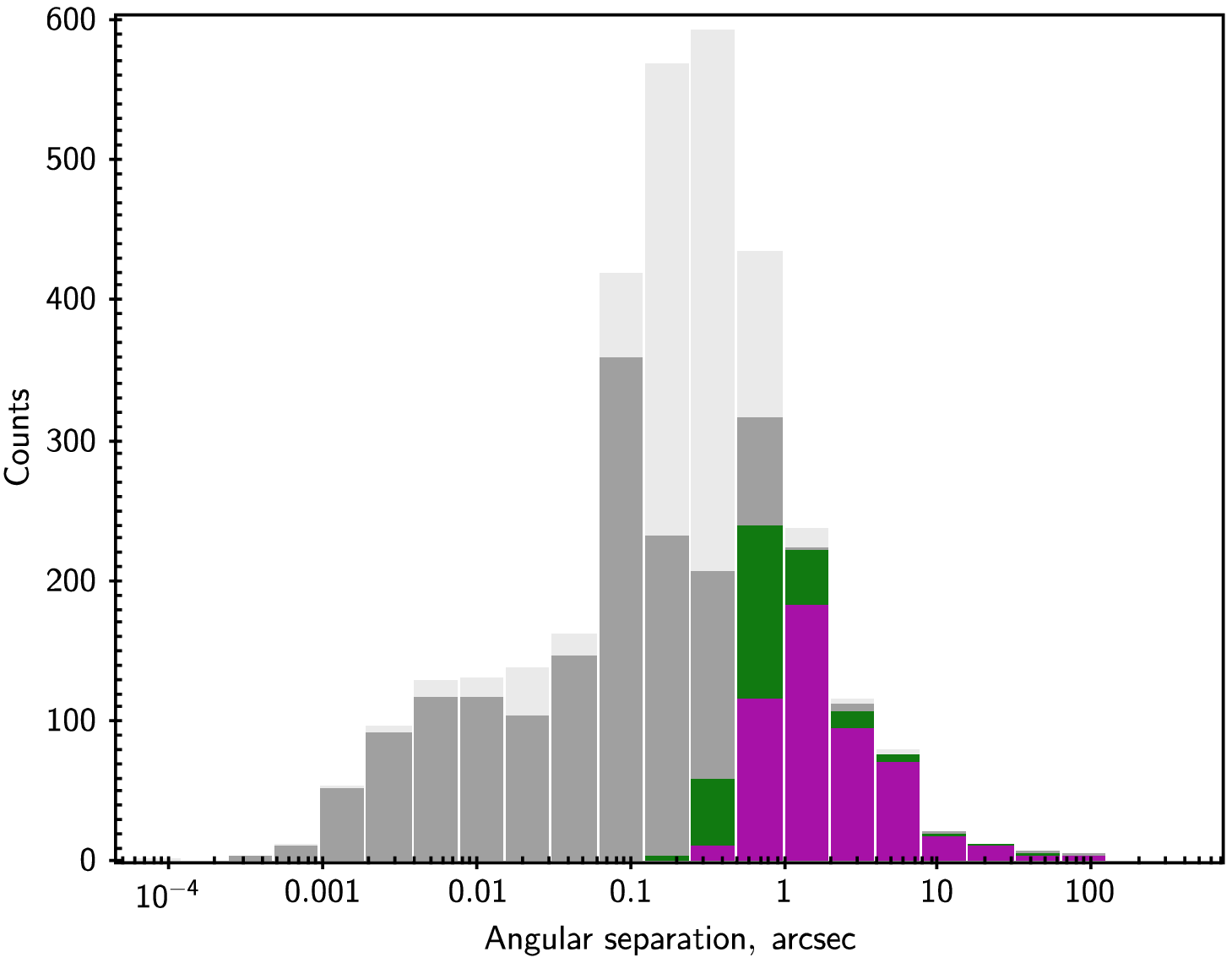}
    \includegraphics[width=\columnwidth]{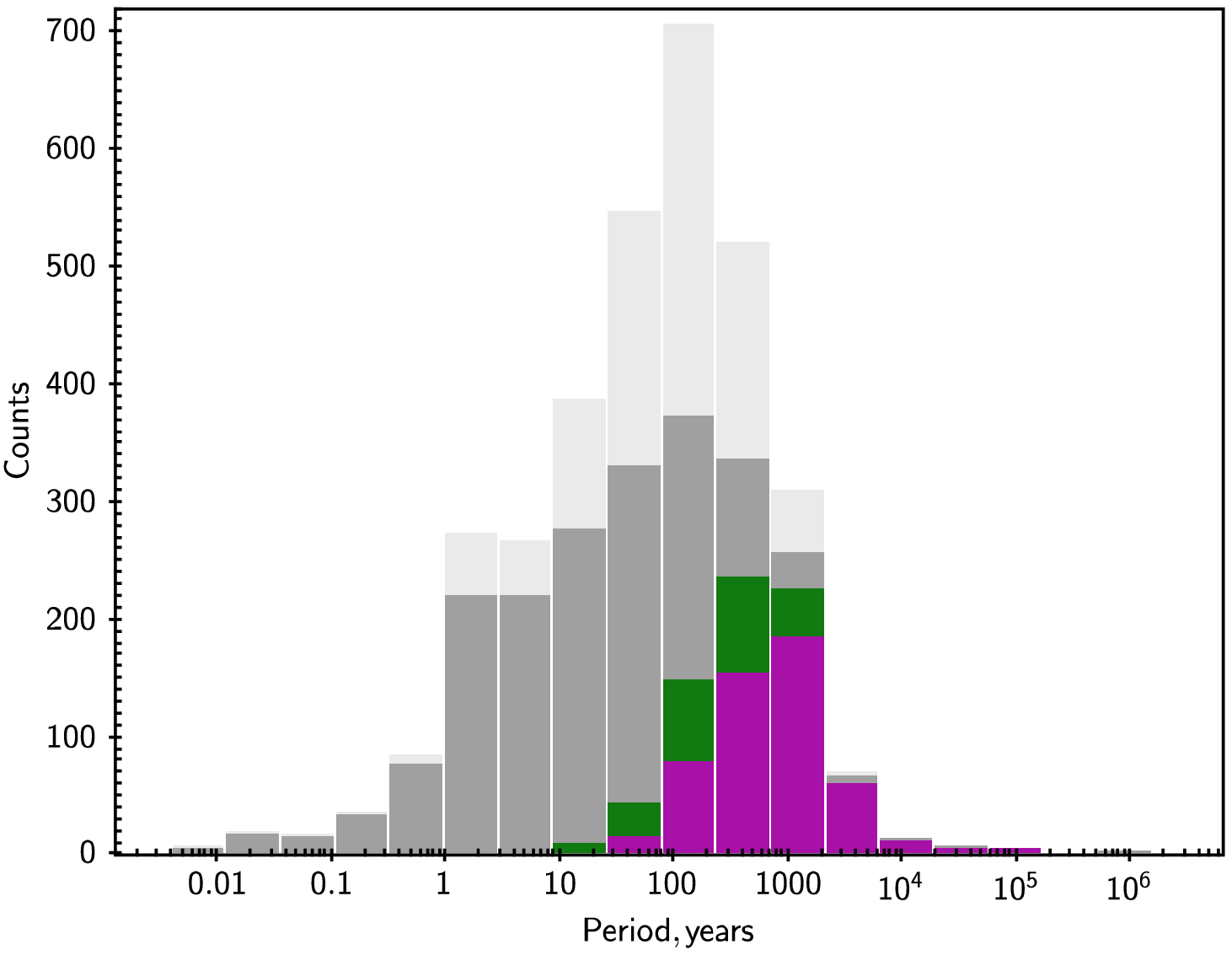}
\caption{Distribution of ORB6 binary stars and their solution type
in EDR3 as a function of ephemeris angular separation and
estimated orbital period. Light gray: pairs with available
solution in \textit{Gaia} EDR3. Dark gray: systems with known EDR3
parallax. Green: resolved pairs. Purple: systems with parallaxes
available for both components.}
    \label{fig:orb6}
    \end{figure}

Depending on angular separation and apparent magnitudes, primary
and secondary components may appear in EDR3 as separate sources;
however, more often only one counterpart is found, see Fig
\ref{fig:orb6}. In such a case, the secondary source is either
essentially merged with the primary or remains undetected due to
large magnitude contrast with respect to the brighter primary. In
multiple systems, the sampled pair can form an unresolved source
with the third component. Notably, in at least
ten cases, the ORB6 binary is the fainter companion of a multiple
system meaning that it is essentially obscured by a brighter star
in close vicinity. With a goal not to lose valuable parallax data,
we choose to accept such identifications and mark them accordingly
in Table \ref{tab:master}. Generally, for the vast majority of
ORB6 entries, EDR3 counterparts are readily identified. Aside from
members of the Sgr $A^*$ cluster, only 26 ORB6 binaries cannot be
matched, 17 of them have primary component mag$_1<2.7$ which is
beyond the bright limit of \textit{Gaia}. Additionally, we did not
find counterparts for 9 pairs with mag$_1 \gtrsim 15$ or unknown
primary magnitude in ORB6. Very faint ORB6 objects are typically
red, brown, or white dwarfs and normally possess high proper
motions due to their proximity to the Sun. Cross-matching of such
objects is challenging, and there is a chance that at least some
of them actually have unidentified 2-parameter solutions in EDR3.

\begin{figure}
    \centering
    \includegraphics[width=\columnwidth]{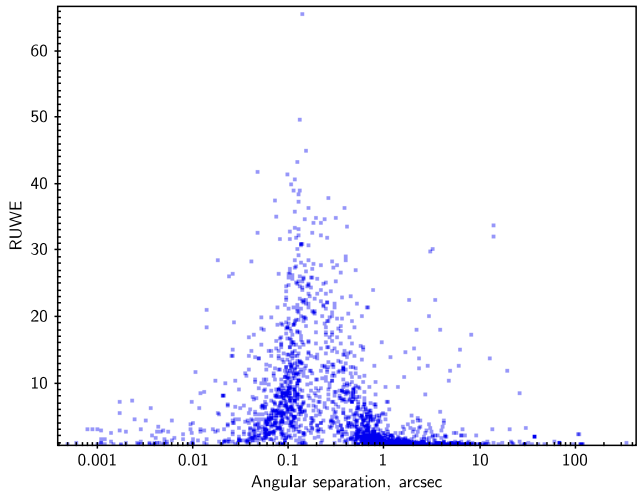}
    \caption{Primary star RUWE as a function of predicted ephemeris separation for ORB6 binaries, orbits with grade 9 are excluded.}
    \label{fig:ruwe}
\end{figure}

Overall for 3279 unique ORB6 entries, at least one \textit{Gaia}
EDR3 counterpart is found, see Table \ref{tab:grades}. 753 pairs
are resolved meaning that their components appear as separate
sources in the EDR3 catalogue. \textit{Gaia} parallaxes are
available for both components in 521 pairs, these systems are
discussed in Section \ref{sec:2plx}. For 1016 entries, we have
2-parameter solutions meaning no parallax available in EDR3. The
solution type heavily depends on the angular separation and
period, see Fig. \ref{fig:orb6}. Clearly, $0.2<\rho<0.5$ arcsec is
the most difficult interval for \textit{Gaia} leaving 2/3 of all
solutions in this range with no parallax. RUWE values for 5- and
6-parameter solutions are high, confirming that these
intermediate-separation binaries do not easily fit the one-star
astrometric solution model of EDR3, see Fig. \ref{fig:ruwe}.
Finally, we mention that 14 \% of primary sources are marked as
duplicated in EDR3. This fraction is higher for binaries with
$0.05<\rho_{\rm eph}<0.2$, reaching 23\%. The fraction of
duplicated sources among secondary stars is just 8\%. For just two
sources, we find the formal EDR3 parallax to be negative, in both
cases the value is comparable to the reported error: WDS
06410+0954 has $\varpi_1=-1.33 \pm 1.00$, and for WDS 17343-1909,
$\varpi_1=-0.02 \pm 0.05$. The former source additionally has
excessive RUWE $\chi_1=5.1$.

\section{Resolved double stars}
\label{sec:2plx}

The identification of resolved binaries is secure, as separation
and positional angle are compared to the values predicted by ORB6
ephemerides. In a few cases, when ORB6 predictions seem
erroneous, WDS data are used as well. The minimum $\rho$ value for
a \textit{Gaia}-resolved pair is 0.20 arcsec; however, in this case,
solutions for both components are 2-parameter ones. Just for 6
resolved binaries with $0.23<\rho<0.35$ arcsec, parallax is
available for one component, while the secondary solution is a
2-parameter one. WDS 00429+2047 has minimal $\rho=0.37$ arcsec
among resolved binaries with two available parallaxes. Notably,
all resolved pairs with $\rho< 0.5$ arcsec have excessive RUWE
$\chi>2.5$.

\subsection{Optical pairs}
\label{sec:optical} ORB6 binaries are expected to be
gravitationally bound. Indeed, the very presence in the ORB6
catalogue is considered {\it a priori} knowledge that the pair is
a physical binary, and strong counter-evidence is needed to refute
the claim. However, often the estimated orbital periods exceed
thousands of years, which is significantly longer than the
observational history. Therefore, a small number of false entries
are anticipated among the ORB6 objects. If an alleged system
appears in EDR3 as a resolved system with parallax and proper
motion available separately for both components, we may
distinguish chance-alignment pairs from genuine physical binaries.

\begin{figure}
    \centering
    \includegraphics[width=\columnwidth]{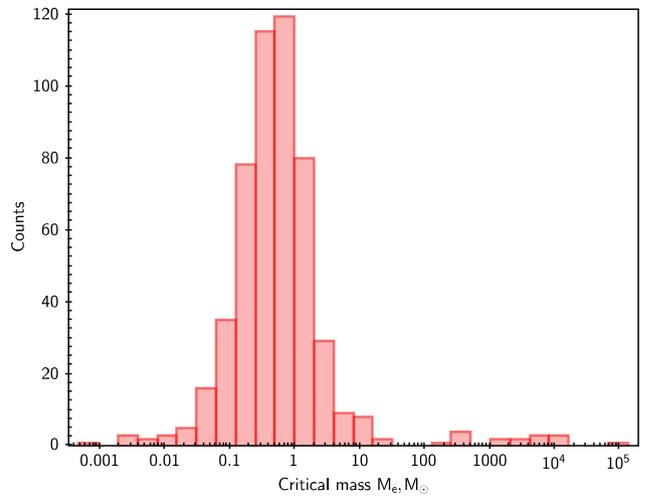}
\caption{Minimal mass required to keep the system gravitationally
bound, see Eq. \ref{eq:escape}. The subset of 521 double stars
with full solutions available for both components is shown. The 16
outlier systems with $M_e>100M_\odot$ are revealed to be likely
optical pairs.}
    \label{fig:min_mass}
\end{figure}

The direct comparison of the components' reported parallaxes
$\varpi_1$ and $\varpi_2$ is possible and discussed in the next
paragraph, but it appears that a more effective way to segregate
optical pairs is to compare proper motions of the components.
Proper motion in right ascension
$\mu_\alpha$\footnote{$\mu_{\alpha}=\mu_{\alpha'} \cdot
\cos(\delta)$, $\mu_{\alpha}$ published in EDR3 is already
corrected for $\cos(\delta)$.} and declination $\mu_\delta$ is
provided for \textit{Gaia} sources with available full solution.
The relative proper motion of components $\Delta \mu =
\sqrt{(\mu_{\alpha 1}-\mu_{\alpha 2})^2+(\mu_{\delta
1}-\mu_{\delta 2})^2}$ can be converted to tangential speed
(measured in km/s; $\varpi$ -- mas; $\Delta \mu$ -- mas
yr$^{-1}$) as:
\begin{equation}
\label{eq:speed}
v \approx \frac{4.74 \cdot \Delta \mu}{\varpi}
\end{equation}
The relative speed of the components should not exceed the escape
velocity $v < \sqrt{2 G_0M/r}$, where $r=\rho/\varpi$ is the
projected distance between the components. The minimum mass
required to keep the system bound is therefore calculated in SI
units as:
\begin{equation}
\label{eq:escape}
M_e=\frac{\rho v^2}{2\varpi G_0}
\end{equation}
$G_0$ is the gravitational constant. As we aim to find the lowest
possible mass, $\varpi$ is chosen as a maximum among components'
parallaxes $\varpi_1$ and $\varpi_2$. The $M_e$ distribution (Fig.
\ref{fig:min_mass}) identifies a small number of outliers with
high required masses, which must be chance alignment systems. For
16 pairs $M_e>100 M_\odot$, while for the rest of the sample,
$M_e<21 M_\odot$. We neglected uncertainties of coordinates and
proper motion reported in \textit{Gaia}. Right ascension and
declination errors are within 1.2 mas being too small to alter
$\rho$. $\mu_{\alpha}$ and $\mu_{\delta}$ uncertainties are more
significant as the relative error reaches 5\% for a few binaries.
Although real errors may be larger than reported, we do not expect
them to change the outcome of optical pairs evaluation.

The comparison of parallaxes $\varpi_1$ and $\varpi_2$ further
suggests that these 16 pairs are optical. As we show in Section
\ref{sec: big_errors}, for a genuine ORB6 binary, the expected
orbit size is negligible in comparison to the naive distance
suggested from parallaxes $|1/\varpi_1-1/\varpi_2|$. If the
measured parallax is distributed according to normal (Gauss) law
with expectation $\varpi_1$, $\varpi_2$ and standard deviation
$\sigma_1$, $\sigma_2$ for the primary and secondary component,
respectively, the difference of parallaxes is expected to follow
Gauss distribution as well. The particular case of equal
parallaxes $\varpi_1=\varpi_2$ has statistical significance:

\begin{equation}
\label{eq:significance}
\frac{\Delta \varpi}{\sigma}=\frac{|\varpi_1-\varpi_2|}{\sqrt{\sigma_1^2+\sigma_2^2}}
\end{equation}

We use $\Delta \varpi /\sigma$ as a measure of parallax agreement
for the components of ORB6 double stars. As we show in the next
section, a formally large parallax disagreement does not
necessarily mean the system is physically unbound. However, all 16
optical pairs candidates show $\Delta \varpi/\sigma > 8$, see
Table \ref{tab:optical}. All suspicious pairs have the
lowest-quality orbits (grade 5), with the exception of WDS
19054+3803 whose orbit is graded 4; therefore to find that these
are chance alignment pairs should not come as a surprise. The
minimal $\rho$ among optical pairs is 2 arcsec in case of WDS
17248+3044.

Finally, one can argue that the critical mass $M_e$ should not
exceed the dynamical mass $M_d$ calculated via Kepler's Third Law,
see Section \ref{sec:dynamical} and Eq \ref{eq:kepler}. Aside from
the 16 likely optical pairs examined above, this condition is
strongly violated by only 3 other systems: WDS 00524-6930, WDS
17419+7209, WDS 19464+3344, which have $\dfrac{M_e}{M_{d}} \sim 3
- 6$. These binaries are characterized by large $\rho>20$ arcsec,
implying that a relatively minor inaccuracy of the proper motion
causes a large deviation of $M_e$. However, their components'
parallaxes show good agreement, $\Delta \varpi / \sigma <1$,
therefore, we do not have sufficient evidence to flag them as
optical pairs.

\begin{table*}
    \centering
\caption{List of 16 ORB6 double stars that are now
suspected to be optical pairs. $g_{1,2}$ are the apparent $G$
magnitudes, $\rho$ is the angular separation. $M_e$ refers to
minimum mass required to keep the system bound, see Eq.
\ref{eq:escape} and Fig. \ref{fig:min_mass}. For each entry,
$M_e>100M_\odot$ and $\Delta \varpi / \sigma > 8$ (Eq.
\ref{eq:significance}). We note that WDS 19127+2435 shows an
exceptionally high RUWE for both components $\chi_1 = 23$ and
$\chi_2 = 30$, which strongly suggests that both sources are
unresolved binaries. Additionally, there is a spurious 2-parameter
source within 0.36 arcsec from the secondary star with a nearly
identical magnitude, thereby, we recommend to be cautious about
this system.}
    \label{tab:optical}
   $ \begin{array}{ccccccccccc}
        \text{Designation} & \multicolumn{2}{c}{\text{Magnitude}}&\text{Grade}&\multicolumn{2}{c}{\text{RUWE}}&\rho& \multicolumn{2}{c}{\text{\textit{Gaia} EDR3 parallax, mas}} & \text{Significance} & M_e  \\
                \hline
\text{WDS}&g_1&g_2 &&\chi_1&\chi_2& \text{arcsec}&\varpi_1 \pm \sigma_1 & \varpi_2 \pm \sigma_2 &  \Delta \varpi / \sigma & M_\odot           \\                       \hline

        00152+2722&11.4&12.1&5&1.0&1.0&6.2& 1.979 \pm 0.025 & 3.522 \pm 0.017 & 50.8 & 3494\\
        03342+4837&7.4&11.2&5&1.0&2.0&5.6&5.936 \pm 0.035&1.633 \pm 0.050& 70.6&491\\
        04599+0031 &10.7&11.2&5&1.0&1.1& 6.0&1.042 \pm 0.016&5.715 \pm 0.021&174.1&525\\
        05013+5015&9.3&9.4&5&0.9&1.0&4.2&4.367 \pm 0.013&4.658\pm 0.014&15.2&310\\
        07106+1543&11.6&12.0&5&1.0&0.9&10.6&12.162 \pm 0.015&2.751\pm 0.014&457&522\\
        08062+0201&10.8&11.1&5&1.0&1.3&2.3&3.911 \pm 0.024&2.114\pm 0.028&49.5&1148\\
        11128+0453&10.1&10.8&5&1.3&1.4&4.1&3.722 \pm 0.023&6.346\pm0.030&69.3&102\\
        17121+2114&7.1&8.8&5&0.9&1.1&8.6&5.256 \pm 0.015&2.304\pm0.015&142&476\\
        17248+3044&10.6&11.0&5&1.7&1.5&2.0&2.398 \pm 0.020&2.163\pm0.021&8.1&360\\
        19054+3803&9.4&9.6&4&0.9&1.0&7.1&1.733 \pm 0.010&1.611\pm0.011&8.4&12593\\
        19127+2435&8.2&11.7&5&22.7&29.9&3.6&13.059 \pm 0.395&1.634\pm0.492&18.1&113\\
        20210+1028&12.2&12.4&5&1.0&0.9&6.8&0.911 \pm 0.011&2.113 \pm 0.011&75.2&81652\\
        21506+2216&8.0&10.8&5&1.0&1.3&2.9&2.200 \pm 0.023&3.544\pm0.027&38.3&443\\
        21559+3141&11.7&11.9&5&1.4&1.7&4.2&2.034 \pm 0.019&2.471\pm0.019&16.6&4143\\
        22280+5742&8.8&12.7&5&1.0&1.1&45.9&0.658 \pm 0.026&0.368\pm0.013&10.0&2780\\
        23100+3651&7.0&7.6&5&1.1&0.9&67.4&2.836 \pm 0.022&1.428\pm0.020&46.5&4258\\

        \hline
    \end{array} $
\end{table*}

\begin{table*}
    \centering
\caption{List of 17 ORB6 visual binary stars with large
disagreement in the reported components' parallaxes $\Delta \varpi
/ \sigma> 8$ (Eq. \ref{eq:significance}), but which are still
believed to be bound systems, see discussion in Section \ref{sec:
big_errors}. The values of orbit quality grade and period are
taken from the ORB6 catalogue.}
    \label{tab:delta_plx}
$ \begin{array}{cccccccccccc}
        \text{Designation}&\multicolumn{2}{c}{\text{Magnitude}}&\text{Grade}&\multicolumn{2}{c}{\text{RUWE}} & \text{Period}&\rho&\multicolumn{2}{c}{\text{\textit{Gaia} EDR3 parallax, mas}} & \text{Significance} & M_e  \\
                        \hline
\text{WDS} &g_1&g_2 &&\chi_1&\chi_2&\text{yr}&\text{arcsec}& \varpi_1 \pm \sigma_1 & \varpi_2 \pm \sigma_2 &\Delta \varpi / \sigma & M_\odot           \\                       \hline
00014+3937&8.9&9.4&3&1.1&1.8&217&1.33&19.34\pm0.02 & 20.03\pm 0.04 & 16.4 & 0.15 \\
00076-0433&9.0&10.0&5&9.6&1.6&688&1.03&17.26\pm0.32&14.15\pm0.05&9.5&0.18\\
03470+4126&7.9&8.5&4&1.1&2.2&2276&6.88&41.90\pm0.02&41.50\pm0.04&8.3&0.48\\
05025-2115&7.8&9.6&3&1.8&1.8&43.6&0.89&119.57\pm0.04&118.82\pm0.08&8.3&0.03\\
06082+3759&7.1&9.6&5&0.9&4.0&9679&1.83&4.54\pm0.03&2.61\pm0.24&8.2&2.62\\
08582+1945&9.2&9.4&4&3.5&2.6&124&2.17&196.26\pm0.20&194.14\pm0.12&9.1&0.12\\
09137+6959&8.7&9.3&5&3.7&15.8&1469&1.15&4.96\pm0.12&7.65\pm0.26&9.3&4.1\\
09551-2632&8.2&8.8&5&4.3&1.5&188&0.86&18.96\pm0.27&23.95\pm0.05&18.0&0.37\\
10217-0946&8.1&10.2&4&3.5&1.1&1340&1.54&15.00\pm0.08&15.71\pm0.03&8.7&0.43\\
10412-3654&7.1&7.6&3&3.2&3.6&60.3&0.84&64.34\pm0.05&62.62\pm0.14&11.2&0.36\\
11214-2027&11.2&11.2&5&18.1&1.4&1028&3.81&72.86\pm0.32&76.19\pm0.03&10.2&0.50\\
11550-5606&4.1&5.1&5&1.9&0.9&972&3.86&31.79\pm0.04&32.23\pm0.02&11.1&0.25\\
12335+0901&4.9&5.6&2&14.0&6.1&15.8&1.15&223.48\pm0.47&231.12\pm0.51&11.0&0.01\\
15348+1032&9.6&9.6&4&3.7&1.2&1150&4.04&10.81\pm0.50&18.97\pm0.11&15.8&0.40\\
16256-2327&7.3&9.3&5&3.4&2.0&4193&3.00&7.26\pm0.13&5.48\pm0.15&8.8&17.1\\
20452-3120&12.5&12.0&5&2.3&2.3&141&2.10&100.79\pm0.07&101.97\pm0.08&11.2&0.40\\
22473-1609&7.3&9.3&5&8.5&1.1&824&2.65&25.49\pm0.20&27.84\pm0.02&11.6&0.34\\
\hline
\end{array}$
\end{table*}

\begin{table}
    \centering
    \caption{The agreement of components' parallaxes (Eq. \ref{eq:significance}) as a function of RUWE.
We consider the set of 505 binaries with full astrometric
solutions available for both componets; optical pairs are excluded
(Section \ref{sec:optical}). Systems are divided into 4 equal-size
groups according to average RUWE of the components,
$\chi=(\chi_1+\chi_2)/2$.  1Q: $\chi \le 1.085$, 2Q: $1.085<\chi <
1.341$, 3Q: $1.341<\chi < 2.37$ , 4Q: $\chi> 2.37$.
    The numbers of pairs in each group with parallax agreement below 5, 3, 2, and 1$\sigma$ are noted; clearly the agreement is better
for stars with low RUWE, see Section \ref{sec: big_errors}.    }
        \label{tab:significance}

    $\begin{array}{ccccc}
    &\multicolumn{4}{c}{\text{RUWE quartile}}\\

     & 1Q & 2Q & 3Q & 4Q \\
                                \hline
\sum \text{stars in quartile} &126&127&126&126\\
\Delta \varpi / \sigma<5 &125&125&112&105\\
\Delta \varpi / \sigma<3 &118&110&93&80\\
\Delta \varpi / \sigma<2 &104&89&67&48\\
\Delta \varpi / \sigma<1 &71&62&47&26\\
\hline
    \end{array}$
\end{table}

\subsection{Binaries with large parallax discrepancy}
\label{sec: big_errors}

After the removal of probable optical pairs, 505 unique binaries
with \textit{Gaia} parallaxes available for both components remain
in the sample. For 206 (41\%) systems, $\varpi_1$ and $\varpi_2$
argee to within one standard error; for 401 (79\%) pairs, the
parallaxes are within 3$\sigma$, see Table \ref{tab:significance}.
While the overall agreement of the parallaxes seems reasonable,
these values are significantly worse than the 68.3 \%  and 99.7 \%
respectively expected from Gaussian statistics. The high number of
outliers with excessive parallax disagreement is particularly
remarkable. Large differences in the  reported parallaxes alone do
not mean that the double star is optical. 17 entries listed in
Table \ref{tab:delta_plx} show adequate critical masses $M_e$ and
therefore are considered to be genuine binaries despite $\Delta
\varpi / \sigma> 8$ which is larger than for some optical pair
candidates. Below we consider some of them in greater detail, as
these examples are rather instructive.

The comparison of parallaxes assumes the orbits are face-on. But
in fact, the orbit of a visual binary star does not generally lie
in the plane of the sky, and consequently, the distances
to the components may differ slightly. This distance difference
will be best measured for systems in the solar vicinity, therefore
we consider WDS 05025-2115, which is a nearby binary with small
reported parallax error, $\sigma_i / \varpi_i < 0.1\%$ for both
components. Naive line-of-sight distance estimate between the
components is $|1/\varpi_1-1/\varpi_2| \approx 0.05 \text{ pc}
\approx 10^4$ AU. This is by far greater than the semi-major axis
estimate for this binary, $a \approx
\frac{a^{\prime\prime}}{\varpi} \approx 10$ AU.
We conclude that the real size of the orbit is
negligible in comparison to distance suggested from $\Delta
\varpi$ and cannot explain the disagreement of parallaxes.

The largest parallax discrepancy in Table \ref{tab:delta_plx}
belongs to WDS 09551-2632 with $\Delta \varpi / \sigma = 18$.
However, we are confident that this system is physical due to its
modest critical mass, $M_e<1.0 M_\odot$. We notice that the
reported $\sigma_1 \gg \sigma_2$. Moreover, this system has an
outer component at $\rho_3=6.2$  arcsec with
$\varpi_3=23.95\pm0.02$, which perfectly matches $\varpi_2$, see
Section \ref{sec:third}. Therefore, we conclude that $\varpi_1$ is
likely to be erroneous.

Unlike optical pairs in Table \ref{tab:optical}, Table
\ref{tab:delta_plx} contains one orbit of grade 2 and three
entries with grade 3, so we are confident that these double stars
are bound based on the historical observations alone. For example,
the second largest $\Delta \varpi / \sigma = 16.4$ belongs to WDS
00014+3937. The values of parallaxes $\varpi_1$ and $\varpi_2$ are
within 3.5\%. However, the reported relative parallax errors are
as small as 0.2\% leading to formally high disagreement. This
system has, however, been observed since 1881, and with an
estimated period 217 years it has a reliable orbit of grade 3.

\subsection{\textit{Gaia} EDR3 uncertainties evaluation}
\label{sec:errors}

The agreement in the parallaxes between the two components of a
system, $ \Delta \varpi/ \sigma$, is strongly correlated with the
parameter RUWE, see Table {\ref{tab:significance}}. We divide the
whole dataset into four equal-size subsamples according to average
RUWE of components $\chi=0.5(\chi_1+\chi_2)$. For $\chi \le 1.085$
subsample, 56\% of parallaxes fit within one $\sigma$ error, which
is rather consistent with Gaussian statistics. The discrepancy is much
worse for the $\chi> 2.37$ subsample, as only 21\% of binaries
meet the $1\sigma$ threshold.

\begin{figure}
    \centering
    \includegraphics[width=\columnwidth]{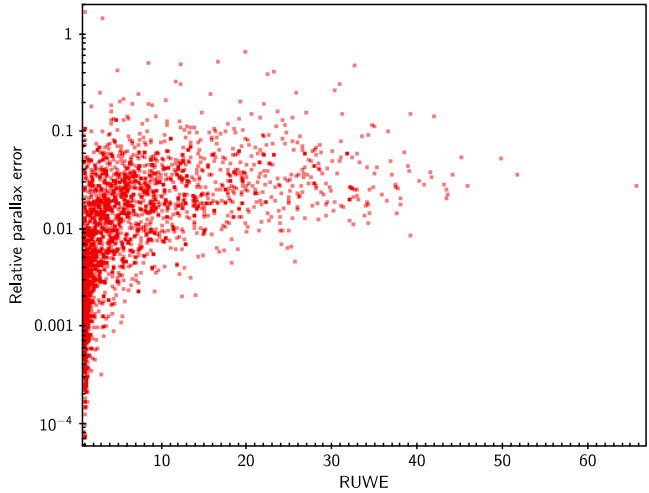}
    \caption{Reported relative parallax error $\sigma_1/\varpi_1$ as a function of RUWE for the primary components of ORB6 pairs.}
    \label{fig:ruwe_error}
\end{figure}

\begin{figure}
    \centering
    \includegraphics[width=\columnwidth]{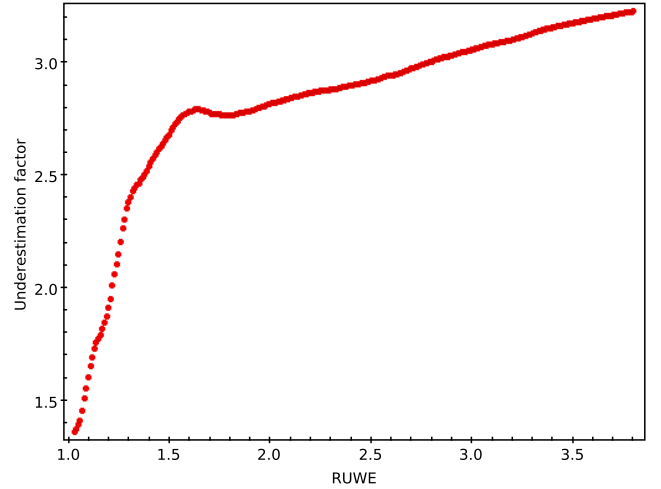}
    \caption{Parallax error underestimation factor $k(\chi)$, see Section \ref{sec:errors}.}
    \label{fig:factor}
\end{figure}

It is remarkable that, while the \textit{Gaia}-reported parallax errors
$\sigma_i/\varpi_i$ for stars with high RUWE are larger (Fig.
\ref{fig:ruwe_error}) than for sources with small $\chi_i$, they
are still lower than what the parallax differences between components
would suggest, which means that the reported $\sigma_i$ values
underestimate the true uncertainties, $\sigma_i^*$. We introduce a
factor $k$, dependent on $\chi$, to evaluate the underestimation
of the reported errors. The formal \textit{Gaia} errors for each
component are multiplied by a factor $k(\chi)$, which is
calculated as $\sigma^*_i=\sigma_i k$, $k>1$, and alternate values
$\Delta\varpi / \sigma^*$ are calculated from Eq.
$\ref{eq:significance}$ for every system. Initially binaries are sorted according
to the average $\chi$ of components.  Next, subsamples of 100
binaries are created from consecutive entries; thus, the first
sample contains the 100 binaries with the lowest RUWE, the next
sample contains binaries from the 2nd to 101st, and the last
\footnote{505-99=406 subsamples are created from the dataset of
505 binaries.} subsample contains the 100 binaries with the
largest $\chi$. The number of binaries with $(\Delta
\varpi/\sigma)_i$<1, and the average RUWE value $\overline\chi$,
are calculated for each subsample. Then, the reported parallax
errors are multiplied by the factor $k$ which is gradually
increased up to 68\% of binaries within the set have
$(\Delta\varpi / \sigma^*)_i < 1$. Thus, we obtain an estimate of the
$k$ factor for a value $\chi = \overline\chi$ from each subsample.
Finally, the obtained values are averaged among 50 neighboring
entries, see Fig. \ref{fig:factor}.

The underestimation factor starts at $k \sim 1.4$ for $\chi \sim
1.05$ and increases at a high rate until it reaches $k \sim 2.7$
at $\chi \sim 1.5$. Then, the relation flattens and shows moderate
growth, reaching $k\sim 3.2$ for $\chi \sim 3.5$. Unfortunately,
due to limited sample size, we are unable to trace $k(\chi)$ for
smaller and larger RUWE. Of course, this correction should be
applied with caution: the correction values $k$ are calibrated
from the one standard error deviation, thus the number of binaries with
high parallax discrepancy will exceed Gauss-distribution
predictions even after this correction.

\section{Third light companions}
\label{sec:third}
1016 ORB6 binaries have a 2-parameter EDR3 solution
which lacks the parallax needed for a mass calculation. The absence of
the desired data prompts us to use indirect methods to obtain
$\varpi$. Many binaries are parts of multiple systems, and their
parallaxes can be retrieved from the outer component which we
refer to as the third light. In certain cases, the parallax of the
third light $\varpi_3$ is more reliable than those of the binary
components, $\varpi_1$ or $\varpi_2$. In Section \ref{sec:
big_errors}, we already introduced outer-companion parallax to
confirm that $\varpi_1$ of WDS 09551-2632 is erroneous and the
secondary component's parallax $\varpi_2$ should be used instead.
However, usually we resort to $\varpi_3$ when both $\varpi_1$ and
$\varpi_2$ are not available.

The crucial problem is to make sure that the third light is an actual companion to the system, and not a chance alignment star. 
Unlike with ORB6 binaries, we cannot rely on historical observations suggesting that the stars
are orbiting round their common centre of mass.
\citet{2022A&A...657A...7K} made a comprehensive search for common
proper motion candidates around \textit{Hipparcos} stars based on
their parallax and tangential velocity. The authors admit that
their list is not exhaustive and misses some bound components due
to rather strict conditions imposed. Our sample is not limited to
\textit{Hipparcos} objects, therefore we choose a more relaxed
approach. We create a set of conditions (Eq. \ref{eq:third_light}
-- \ref{eq:third_light2}) intended to remove chance-alignment
stars and apply it to every ORB6 entry. Third-light companions
around binaries with known full solutions are used for validation
of equations, which are restricted until no suspected unbound
components make the cut. WDS provides proper motion $\mu_W$ for
most of the ORB6 entries which we use as the reference and test
the relation of the third light with a binary calculating the
critical mass $M_e$ with Eq. $\ref{eq:speed}$  and
$\ref{eq:escape}$, adopting $\varpi=\varpi_3$ and $\Delta \mu =
\sqrt{(\mu_{\alpha_ {W}}-\mu_{\alpha
3})^2+(\mu_{\delta_{W}}-\mu_{\delta 3})^2}=\sqrt{\Delta
\mu^2_\alpha + \Delta \mu^2_\delta}$. For binaries with available
full solutions, $M_e$ is additionally calculated according to the
relative proper motion of the primary star $\mu_1$ and the third
light: $\Delta \mu = \sqrt{(\mu_{\alpha 1}-\mu_{\alpha
3})^2+(\mu_{\delta 1}-\mu_{\delta 3})^2}$. The latter is important
as \textit{Gaia}'s $\mu$ data are expected to be more reliable
than those of the WDS. For a random chance-alignment star, the
critical mass is unrealistically large, therefore we constrain
conditions on $\Delta \mu$ until stars with excessive $M_e$ are
gone. This procedure works fine for binaries with high $\mu$, as
their proper motions stand out among field stars. The larger
precision is required for slower stars. We suspect that $M_e$ is
inadequately calculated for binaries with large RUWE reflecting
the poor quality of the solution, therefore we are more tolerant
to excessive values for such systems. The risk of contamination
from field stars is low for close and bright components, therefore
we add them without additional assumptions on proper motion. The
receipt in Eq. \ref{eq:third_light} is adopted, $\mu$ is measured
in mas yr$^{-1}$, $\rho_3$ is the angular distance between the
third-light companion and the reference ORB6 coordinates. In the
latter expression, the primary star EDR3 coordinates $\alpha_1$,
$\delta_1$ are used for $\overline{\rho_3}$ calculation, $g$ is
the apparent magnitude of the third light.

\begin{equation}
\label{eq:third_light}
\begin{cases}

    |\mu|>250 ,  \Delta \mu_{\alpha,\delta} <125\text{ , } \rho_3<180^{ \prime \prime}\\
    |\mu|>200 , \Delta \mu_{\alpha,\delta} < 50\text{ , } \rho_3<180^{ \prime \prime}\\
    |\mu|>105 , \Delta \mu_{\alpha,\delta} < 15\text{ , } \rho_3<180^{ \prime \prime}\\
    |\mu|>55 , \Delta \mu_{\alpha,\delta} < 5\text{ , } \rho_3<180^{ \prime \prime}\\
    |\mu|>90 , \Delta \mu_{\alpha,\delta} < 25\text{ , } \rho_3<90^{ \prime \prime}\\
    |\mu|>30 , \Delta \mu_{\alpha,\delta} < 5\text{ , } \rho_3<30^{ \prime \prime}\\
    \overline{\rho_3}<8^{ \prime \prime} \text{ , } $g<10$

\end{cases}
\end{equation}

\begin{figure}
    \centering
    \includegraphics[width=\columnwidth]{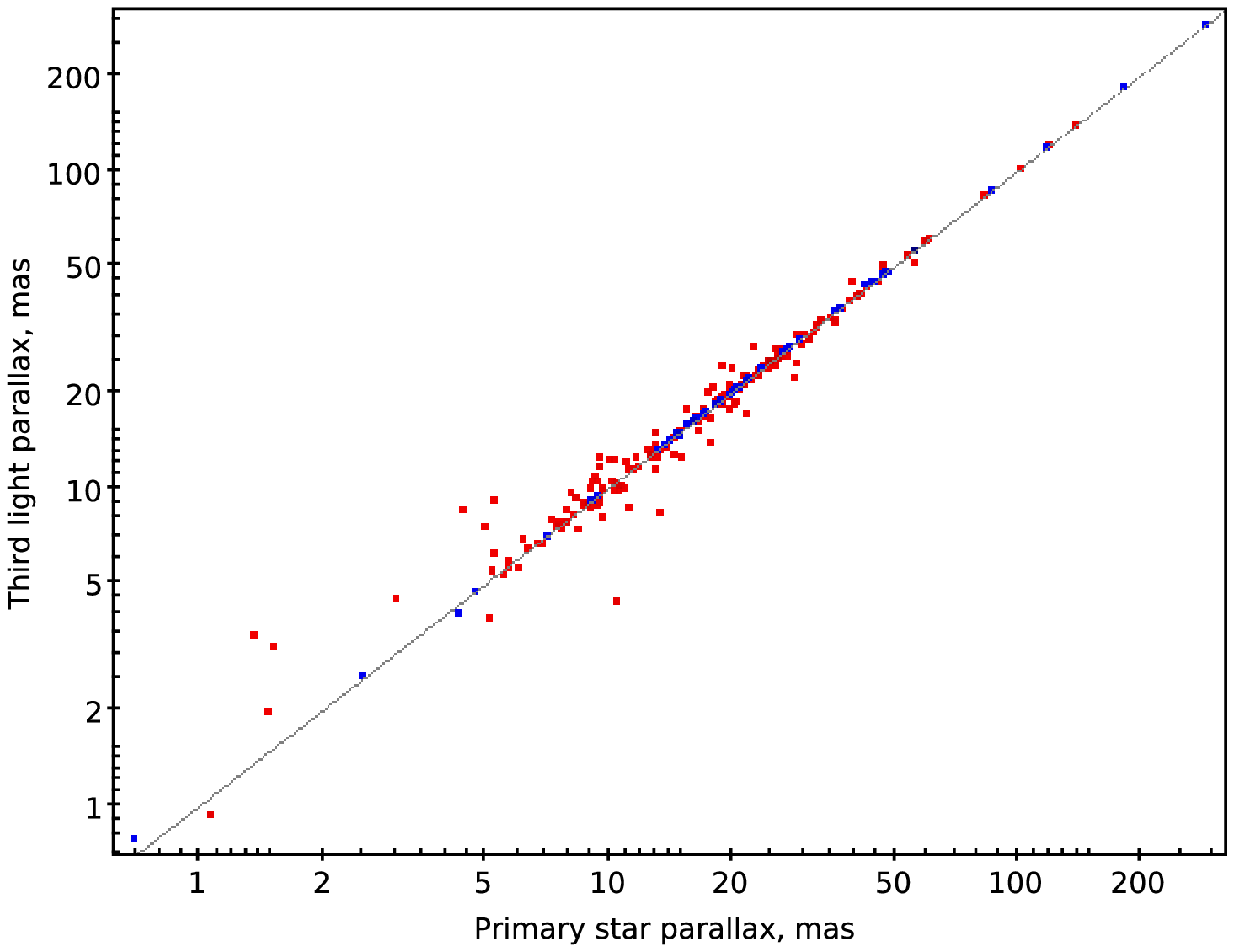}
\caption{Parallaxes of the third-light companions processed
through Eq. \ref{eq:third_light} compared to parallaxes of ORB6
primary stars. Blue dot is used when both sources have reliable
low-RUWE ($\chi_{1,3}<1.8$) solutions. Sources with larger $\chi$
(coloured red) show a larger scatter. The straight line stands for
$\varpi_1=\varpi_3$.}
    \label{fig:plx1_plx3}
\end{figure}

The comparison of parallaxes for the third light and the primary star
serves as validation, see Fig. \ref{fig:plx1_plx3}.
Overall, 380 unique third-light components are
found with Eq. \ref{eq:third_light}, 130 of them are related to
binaries which lack both $\varpi_1$ and $\varpi_2$. The number of
known third-light components can be increased if we introduce
a reference parallax $\varpi_R$ as \textit{a priori} information
to include those with moderate or low proper motion.
Similarly to Eq. \ref{eq:third_light}, conditions are carefully
adjusted to reveal more genuine companions leaving aside the
chance alignment stars. Although the $\varpi_3$ values are close
to the previous $\varpi_R$ ones, the former have better accuracy
and provide an improvement for a system.

\begin{equation}
\label{eq:third_light2}
\begin{cases}

 \varpi_{R}>10, \dfrac{2\varpi_R}{3}<\varpi_3<\dfrac{3\varpi_R}{2}\text{ , }  \Delta \mu_{\alpha,\delta} < 10  \text{ , }  \rho_3<180^{ \prime \prime}\\

 \varpi_{R}>10, \dfrac{\varpi_R}{2}<\varpi_3<2\varpi_{R} \text{ , }  \Delta \mu_{\alpha,\delta} < 50  \text{ , }  \rho_3<30^{ \prime \prime}, $g<19.5$\\

 \varpi_{R}>5, \dfrac{4\varpi_R}{5}<\varpi_3<\dfrac{5\varpi_{R}}{4}\text{,}  \Delta \mu_{\alpha,\delta} < 25  \text{ , }  \rho_3<100^{ \prime \prime} , $g<12$\\

\varpi_{R}>2, \dfrac{4\varpi_R}{5}<\varpi_3<\dfrac{5\varpi_{R}}{4}\text{,}  \Delta \mu_{\alpha,\delta} < 10  ,  \rho_3<60^{ \prime \prime} , $g<12$\\

\end{cases}
\end{equation}

\begin{figure}
    \centering
    \includegraphics[width=\columnwidth]{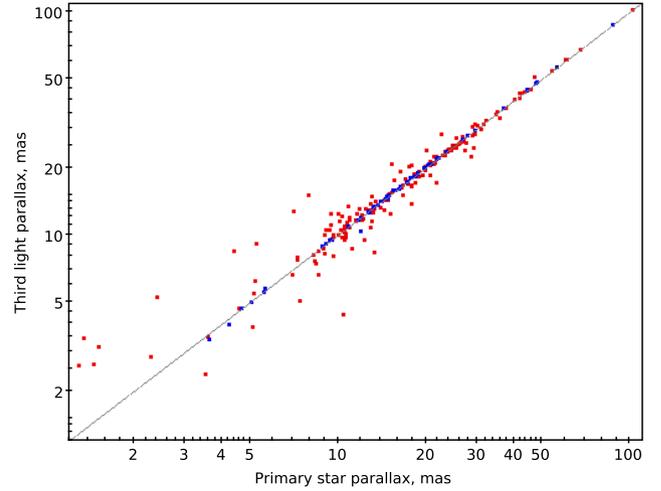}
\caption{Parallaxes of the third-light companions processed
through Eq. \ref{eq:third_light2} compared to parallaxes of ORB6
primary stars. \textit{Hipparcos} parallax
\citep{2007ASSL..350.....V} serves as reference $\varpi_R$. Blue
dot is used when both sources have reliable low-RUWE
($\chi_{1,3}<1.8$) solutions. Sources with larger $\chi$ (coloured
red) show a larger scatter. The straight line stands for
$\varpi_1=\varpi_3$.}
    \label{fig:plx1_plx3_2}
\end{figure}

The latter equation alone allows to reveal 384 or
282 third-light companions when either the \textit{Hipparcos}
parallax $\varpi_{H07}$ \citep{2007ASSL..350.....V} or the EDR3 parallax $\varpi_1$ are used as the
reference $\varpi_R$. It should be reminded that $\varpi_1$ is
unavailable for almost 1/3 of the sample objects, therefore, the
use of \textit{Hipparcos} parallaxes is more fruitful.
In 28 cases, the third-light companion
is in a significant disagreement with the reported parallax of the
close pair, see Fig. \ref{fig:plx1_plx3_2} and Table
\ref{tab:third_light}. If several objects are found with Eq.
\ref{eq:third_light} -- \ref{eq:third_light2}, that with the
largest $\varpi_3/\sigma_3$ is chosen. We reject the high-RUWE
($\chi_3>1.8$) third-light source when a more reliable
($\chi_3>1.5\chi_1$) primary parallax exists because it can be
misleading. Anyway, caution should be exercised when using
$\varpi_3$. It is always possible that the third light is a part of
a co-moving stellar group, meaning that the line-of-sight
distance can be significantly larger than the projected distance.
Still we believe that these data can be helpful, especially when
other sources of parallaxes are non-existent. For nine more ORB6
entries with absent parallax data (see the next section), we
identified probable third-light companions that do not strictly
meet the conditions in Eq. $\ref{eq:third_light}$. Overall, 548 unique third-light companions are found and 196 ORB6 entries with 2-parameter EDR3 solution  are supplied
with EDR3 parallax from the third-light component.

\begin{table*}
    \centering
\caption{List of ORB6 binaries which have a third-light companion
whose reported parallax differs significantly from that of the binary
members, see discussion in Section \ref{sec:third} and Eq.
\ref{eq:third_light}--\ref{eq:third_light2}.   Systems with
$\frac{|\varpi-\varpi_3|}{\sqrt{\sigma^2+\sigma_3^2}} > 8$,
$\varpi/\varpi_3<0.8$, or $\varpi_3/\varpi<0.8$ are shown.
$\varpi$ is the weighted arithmetic mean of the parallax of the
binary components (Eq. \ref{eq:arithmetic}.), however, for the
selected binaries $\varpi_2$ is not available, so $\varpi_1$ is
used in its place. With the exception of WDS 17082-0105, all
listed binaries have an unconvincing high-RUWE solution for the
primary star which is refuted by the more accurate third-light
parallax. \textit{Hipparcos} parallax $\varpi_{H07}$ \citep{2007ASSL..350.....V} is shown for the
reference. We recommend to consider the use of $\varpi_3$ instead of
$\varpi_1$ for the listed systems.}

    \label{tab:third_light}
$    \begin{array}{ccccccccc}

    \text{Designation} & \multicolumn{3}{c}{\text{Parallax, mas}} & \text{Significance}& \multicolumn{2}{c}{\text{RUWE}} & \text{Separation}& \text{Linear separation}\\
        \hline
        \text{WDS} & \varpi_1 \pm \sigma_1& \varpi_3 \pm \sigma_3 & \varpi_{H07} & \frac{|\varpi_1-\varpi_3|}{\sqrt{\sigma_1^2+\sigma_3^2}} &\chi_1&\chi_3 & \rho_3, \text{arcsec} & \rho_3/\varpi_3, 10^3 AU\\
                \hline
00046+4206 & 3.00 \pm 0.35 & 4.43 \pm 0.02 & 2.59 \pm 0.56 & 4.0 & 10.7 & 1.2 & 5.3 & 1.21 \\
00090-5400 & 4.95 \pm 0.42 & 7.49 \pm 0.07 & 7.79 \pm 0.74 & 6.0 & 19.8 & 1.0 & 18.2 & 2.43 \\
01263-0440 & 21.78 \pm 0.54 & 17.01 \pm 0.06 & 18.88 \pm 1.98 & 8.8 & 18.3 & 1.1 & 24.6 & 1.45\\
02529+5300 & 5.09 \pm 0.42 & 3.85 \pm 0.05 &3.25 \pm 24.55 & 2.9 & 11.1 & 1.1 & 1.6 & 0.407\\
03184-2231 & 7.08 \pm 1.47 & 12.63 \pm 0.02 & 12.77 \pm 1.17 & 3.8 &10.5&1.1& 29.4 & 2.33 \\
04316+1743 & 28.51 \pm 0.58 & 22.24 \pm 0.02 & 22.76 \pm 1.21 & 10.8&43.3&1.2&119&5.36\\
05133+0252 & 13.33 \pm 0.31 & 8.30 \pm 0.04 & 9.32 \pm 0.94& 16.1& 4.3 & 1.0&6.9&0.83 \\
05182+3322 & 9.51 \pm 0.57 & 12.45 \pm 0.04 & 14.04 \pm 0.58 & 5.2& 5.4 & 1.1 & 4.2 & 0.335 \\
05484+2052 & 7.39 \pm 0.36 & 5.06 \pm 0.03 & 5.97 \pm 0.73 & 6.5 & 6.2 & 1.3 & 75.4 & 14.9 \\
06410+0954 & -1.33 \pm 1.00 & 1.40 \pm 0.10 & 3.55 \pm 0.5 & 2.7 & 5.1 & 1.3 & 3.0 & 2.12\\
06594+2514 & 2.42 \pm 0.61 & 5.26 \pm 0.03 & 4.48 \pm 2.89 & 4.7 & 25.8 & 1.8 & 22.2 & 4.22 \\
08291-4756 & 1.48 \pm 0.24 & 1.95 \pm 0.03 & 1.53 \pm 0.34 & 1.9 & 4.2 & 0.9 & 3.5 & 1.78\\
10223+4130 & 17.80 \pm 0.39 & 13.77 \pm 0.28 & 14.16 \pm 0.54 & 8.4 & 3.3 & 0.9 & 55.2 &4.01 \\
11151+3735 & 1.35 \pm 0.41 & 3.44 \pm 0.11 & 3.18 \pm 1.17 & 4.9 & 12.2 & 1.7 & 0.6 & 0.183\\
11551+4629 & 10.44 \pm 0.66 & 4.39 \pm 0.02 & 4.72 \pm 0.58 & 9.1 & 19.0 & 1.0 & 3.9 & 0.892\\
12064-6543 & 8.61 \pm 0.74 & 6.60 \pm 0.02 & 7.94 \pm 4.49 & 2.7 & 24.1 & 1.0 & 8.8 & 1.33\\
13099-0532 & 11.18 \pm 0.41 & 8.67 \pm 0.02 & 10.33 \pm 1.09 & 6.1 & 2.4 & 1.0 & 7.0 & 0.809 \\
13123-5955 & 5.26\pm 0.82 & 9.10 \pm 0.15 & 8.61 \pm 0.85 & 4.6 & 14.7 & 3.1 & 1.9 & 0.209\\
15071-0217 & 12.31 \pm 0.81 & 9.47 \pm 0.13 & 10.50 \pm 0.83 & 3.5 & 25.5 & 0.7 & 7.9 & 0.833 \\
16073-3645 & 15.29 \pm 0.67 & 20.63 \pm 0.03 & 18.25 \pm 1.05 & 7.9 & 14.3 & 1.5 & 40.4 & 1.96\\
17082-0105 & 11.91 \pm 0.09 & 10.35 \pm 0.13 & 11.17 \pm 0.95 &9.7& 1.5 & 1.0 & 39.6 & 3.82\\
18232-6130 & 4.40 \pm 0.89 & 8.43 \pm 0.05 & 6.96 \pm 1.03 & 4.5 & 5.2 & 1.2 & 3.6 & 0.425\\
18421+3445 & 1.30 \pm 0.67 & 2.61 \pm 0.03 & 2.19 \pm 0.61 & 1.9 & 16.6 & 1.0 & 25.0 & 9.61\\
19431-0818 & 7.96 \pm 0.50 & 15.02 \pm 0.23 & 13.85 \pm 0.63 & 12.7 & 30.0 & 8.8 & 96.6 & 6.43\\
19579+4216 & 1.52 \pm 0.59 & 3.13 \pm 0.02 & 3.07 \pm 0.47 & 2.7 & 22.4 & 1.0 & 3.0 & 0.943 \\
20169+5017 & 22.51 \pm 0.41 & 28.00 \pm 0.01 & 32.50 \pm 0.55 & 13.3 & 24.8 & 1.1 & 106 & 3.78\\
20593+1534 & 3.54 \pm 0.21 & 2.37 \pm 0.02 & 2.45 \pm 1.18 & 5.5 & 7.2 & 1.1 & 34.4 & 14.5 \\
20598+4731 & 1.47 \pm 0.37 & 2.63 \pm 0.01 & 2.3 \pm 0.42 & 3.2 & 2.7 & 0.8 & 20.3 & 7.72\\

\hline
    \end{array}$
\end{table*}

\section{Other sources of parallaxes}
\label{sec:outer}
889 unique ORB6 pairs are left without EDR3
parallax. Below we briefly discuss data retrieved for sample objects
from other catalogues. Their proper analysis goes beyond the scope
of this paper, here we restrict ourselves to several necessary
remarks. We start with the predecessor \textit{Gaia} DR2 catalogue
\citep{2018A&A...616A...1G}. Around 35\% of the sample sources
with 2-parameter solution in EDR3 have parallaxes available in
DR2. The DR2 parallaxes are based on a shorter data-collection
time, and the absence of EDR3 parallax  automatically means that
the DR2 parallax for a given source is potentially wrong. Still,
considering that for some binaries it is the only available data
source, we choose not to ignore it. Search of DR2 counterparts is
based on EDR3 coordinates and is rather straightforward due to the
small half-a-year difference of coordinate epochs. For more than
two thousand systems, both DR2 and EDR3 parallaxes are jointly
available. We use Eq. \ref{eq:significance} as a measure of
parallax disagreement with $\varpi_1$ and $\varpi_2$ standing for
the reported parallaxes of primary stars in DR2 and EDR3. For 32\%
of primary stars, $\Delta \varpi / \sigma < 1$, and for 73\%,
$\Delta \varpi / \sigma < 3$. A large number of stars show
enormous differences between DR2 and EDR3 parallaxes, and we
expect that many DR2 parallaxes for sources with 2-parameter EDR3
solution are peculiar and therefore should be applied with
caution.

The \textit{Hipparcos} catalogue \citep {1997A&A...323L..49P},
along with its revised \citep{2007ASSL..350.....V} version,
provides parallaxes for most of the ORB6 binaries. Owing to
dedicated solutions accounting for the non-linear motion in
multiple systems, we may expect that, in certain cases, the
\textit{Hipparcos} parallaxes are more reliable than those from
\textit{Gaia} EDR3. Our sample is mostly comprised of relatively
bright sources which limits \textit{Gaia}'s superiority, and the
reported \textit{Hipparcos} relative parallax errors $\sigma /
\varpi$ are just half an order of magnitude larger on average.
Comparing to EDR3, 59\% of parallaxes meet one standard error (Eq.
\ref{eq:significance}) for the original catalogue and 47\% for the
revised version. The most discrepant parallaxes show large RUWE in
EDR3. Two binaries (WDS 02382+4604 and
16229+3803) with no EDR3, DR2, or \textit{Hipparcos} parallax are
found in the \textit{Tycho}-\textit{Gaia} Astrometric Solution
catalogue \citep{2016A&A...595A...2G}. For the remaining 82 entries without available $\varpi$, we conduct a search in the SIMBAD database \citep{2000A&AS..143....9W} to
obtain distance or parallax estimates for all remaining systems,
the related references are provided in Table \ref{tab:master} of
the Appendix.

\section{Mass estimation methods}
\label{sec:masses}

\subsection{Dynamical mass}
\label{sec:dynamical} As mentioned in the introduction, Kepler's Third Law allows one to calculate the mass of a visual
binary from the known orbital period and size. The derived value
represents the total inside the binary's orbit, for a proper
binary it is just the sum of the masses of the primary and
secondary. However, the value potentially includes contribution
from unseen components.

\begin{equation}
\label{eq:kepler}
M_{d}=\frac{a^3}{P^2}=\frac{a^{\prime\prime3}}{\varpi^3 P^2},
\end{equation}
where the mass is measured in $M_\odot$ if the semi-major axis
$a^{\prime\prime}$ and parallax $\varpi$ are in the same angular
units, and the period $P$ is in years. When the parallaxes
$\varpi_{1,2}$ of both companions are available, a weighted
arithmetic mean is used:
\begin{equation}
\label{eq:arithmetic}
k_i=1/\sigma_i^2 \text{ ; } \varpi=\frac{k_1 \varpi_1 + k_2 \varpi_2}{k_1+k_2} \text{ ; } \sigma=\frac{1}{\sqrt{k_1+k_2}}
\end{equation}

For the majority of ORB6 binaries, all three contributing
parameters in Eq. \ref{eq:kepler} are known along with relatively large
measurement errors, see Table \ref{tab:grades}.  Due to error
propagation, the relative uncertainties of orbit size and parallax
are essentially multiplied by a factor of 3, period errors are
doubled, e.g. 10\% error of $a^{\prime\prime}$ turns into $\sim
30\%$ $M_{d}$ uncertainty. Data on $P$ and $a^{\prime\prime}$ are
obtained from diverse methods and therefore are highly
inhomogeneous. In fact, $P$ and $a^{\prime\prime}$ are derived
concurrently from orbital solutions, which means that their
best-fit values are often correlated. Moreover, some methods use
parallax for the solution of visual binary orbits
\citep{2017AstL...43..316K}. These factors impede uncertainty
analysis and require comprehensive evaluation for each system,
which is beyond our capability. We take a simplified approach to
give a rough error estimate for $M_{d}$ and assume all parameters
to be independent. We randomly generate $10^6$ values of $P$,
$a^{\prime\prime}$, and $\varpi$ assuming Gauss distributions for
uncertainties of the reported parameters. When $P$ or
$a^{\prime\prime}$ errors are absent for a given orbit in ORB6,
the 75\% quantile for the respective orbit quality grade is used.
The median of the resulting sample is believed to represent the
average value of $M_d$, while the 0.159 and 0.841 quantiles are
used as confidence intervals $M_d^-$ and $M_d^+$. This approach is
questionable for binaries with poorly defined orbital parameters
typical of grades 4--5, and their formal error estimates and
median masses are of poor credibility, therefore we also provide
the value of Eq. \ref{eq:kepler}   $M_d^{0}$ which completely
ignores the uncertainties.

\subsection{Mass -- luminosity relation}
\label{sec:mlr} The mass--luminosity relation (MLR) was originally
discovered following observations of visual binary stars
\citep{1923BAN.....2...15H,1923PASP...35..189R} and can be used to
estimate masses of main-sequence (MS) stars with known parallax
and brightness. MLRs available from the literature are normally
expressed in terms of bolometric or Johnson $V$ magnitudes, and
their conversion to the \textit{Gaia} photometric system is not
trivial as it depends on the stellar type. Particularly, the $G-V$
relation suggested by the \textit{Gaia} team is invalid for red M
dwarfs with $B-V>1.3$ \citep{2021A&A...649A...3R}. We attempt to
derive the MLR in the EDR3 photometric system using MIST (version 1.2) 
evolutionary tracks \citep{2016ApJ...823..102C,
2016ApJS..222....8D} and PARSEC (v1.2S release) isochrones
\citep{2012MNRAS.427..127B,2014MNRAS.444.2525C,2015MNRAS.452.1068C}. The desired MLR should provide
stellar mass as a function of absolute $G$ magnitude. Stars of
various evolutionary status and metallicity populate the MS,
creating substantial dispersion in the MLR
\citep{2021MNRAS.507.3583E,2021A&A...647A..90F}. Since solar
metallicity  $[M/H]=0$ is typical of stars in the solar
neighbourhood, which our sample belongs to, we adopt it for
further calculations.

\begin{figure}
    \centering
    \includegraphics[width=\columnwidth]{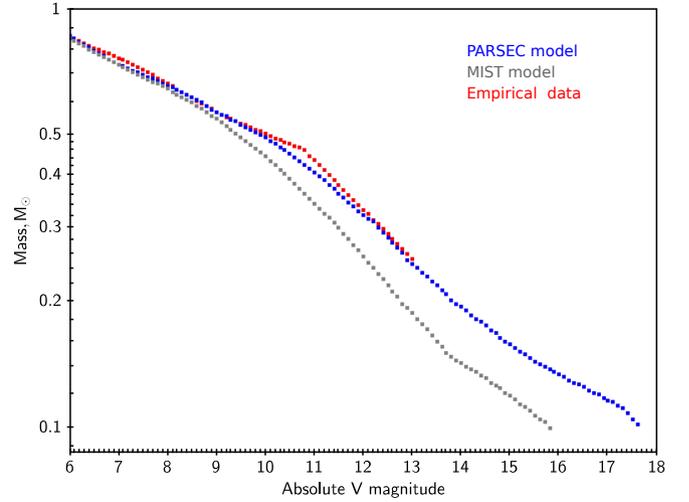}
\caption{Comparison of synthetic ZAMS from the MIST (grey) and
PARSEC (blue) models for low-mass stars, and interpolated
empirical data by \citet{2020MNRAS.496.3887E} (red). The PARSEC
model better fits the empirical data for the faint stars ($V>8$
mag).}
    \label{fig:zams}
\end{figure}

\begin{figure}
    \centering
    \includegraphics[width=\columnwidth]{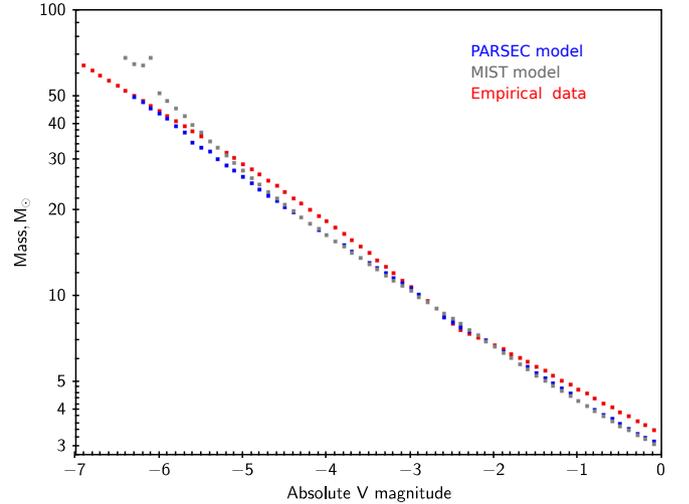}
\caption{Comparison of synthetic IAMS from the MIST (grey) and
PARSEC (blue) models for high-mass stars, and interpolated
empirical data by \citet{2020MNRAS.496.3887E} (red). While both
models are nearly identical, for $V<-5$ mag, the PARSEC model is
clearly preferable.}
    \label{fig:iams}
\end{figure}

The MIST model provides absolute magnitude as a function of age
for a star with a given initial mass. A dense grid of masses in
$-1 <\log{M/M_\odot} < 2$ range with 0.01 dex step is adopted.
Conveniently, the model explicitly defines zero-age and
intermediate-age main-sequence (ZAMS and IAMS) evolutionary
phases, therefore we acquire synthetic ZAMS and IAMS provided with
stellar age, mass, and absolute $V$ and $G$ magnitudes. Massive
stars have a relatively short lifespan; assuming smooth star
formation history, their mean luminosity is expected to be
represented by middle-age stars close to the IAMS\footnote{IAMS is
defined as an evolutionary phase with 30\% mass fraction of
hydrogen in the stellar centre \citep{2016ApJS..222....8D}.}.
Low-mass stars, which have a total lifespan comparable to or
larger than the Hubble time, should be close to the ZAMS. These
basic considerations are confirmed by the comparison to empirical
data, see Fig. \ref{fig:zams} and \ref{fig:iams}.

Using MIST ages and masses as input, we additionally calculate
synthetic ZAMS and IAMS with the PARSEC isochrones. For this
purpose, the logarithm of age is rounded to 0.01 dex precision,
while absolute magnitudes are linearly interpolated as a function
of mass logarithm. Thus we obtain dense synthetic ZAMS and IAMS
grids for both MIST and PARSEC models which provide stellar mass
as a function of absolute $V$ and $G$ magnitudes. We notice that
synthetic ZAMS are generally consistent in both models, as
predicted masses are within 2\% for stars with $G<8.5$ ($M \gtrsim
0.6 M_\odot$). However, for lower masses the MIST and PARSEC
models significantly diverge, see Fig. \ref{fig:zams}. In a
similar way, synthetic IAMS are consistent over a wide mass range
but diverge at the highest masses, see Fig. \ref{fig:iams}. For
validation, we use table~6 from \citet{2020MNRAS.496.3887E}, which
is based on empirical relations and provides absolute $V$
magnitudes for MS stars in the 0.25 $M_\odot$ -- 64 $M_\odot$
range. The PARSEC model clearly shows better agreement with
empirical data; therefore we adopt it for our calculations,
leaving MIST as an evolutionary phase marker.

Since neither the ZAMS or IAMS alone can reproduce the empirical
data for each evolutionary track, we interpolate between the ZAMS
and IAMS in ten equal-age steps to obtain grids of synthetic
populations with identical evolutionary phase (isophase). Next,
for any given $V$, we select the grid line showing the best
agreement of synthetic and empirical masses. As expected, the
low-mass stars are better approximated by the ZAMS, while for
large masses, grids with advanced evolution are preferred, and the
most massive stars are best-fitted by IAMS. The transition is not
completely monotonous, reflecting the limitations of synthetic
tracks and empirical data. Thus, we create a composite synthetic
grid closely resembling empirical MLR in the $V$ band and apply
the same isophase grids to get the $G$-band MLR. The resulting MLR
for $V$ and $G$ magnitudes is shown in Fig. \ref{fig:mlr} and is
available in Table \ref{tab:mlr} of the Appendix. Stars with
$M<0.1 M_\odot$ are not described by synthetic models, therefore,
for the $0.075 M_\odot<M<0.1 M_\odot$ range we use the updated
\citet{2013ApJS..208....9P} table based on
\citep{2019MNRAS.485.4423S} data. The agreement between synthetic
and empirical MLR is within 0.1 mag in $V$-band for $M>0.8
M_\odot$ and increases to 0.25 mag at the lowest masses. The
intrinsic MLR uncertainty depends on stellar mass and does not
exceed 38\% \citep{2018MNRAS.479.5491E,2021MNRAS.507.3583E} which
translates into 0.35 magnitude error; however, the dispersion is
approximately two times smaller for low-mass stars. Considering
errors introduced by conversion to $G$ magnitudes and intrinsic
MLR dispersion, we assume a general $\sigma_{\text{MLR}}=0.4$ mag
uncertainty.

\begin{figure}
    \centering
    \includegraphics[width=\columnwidth]{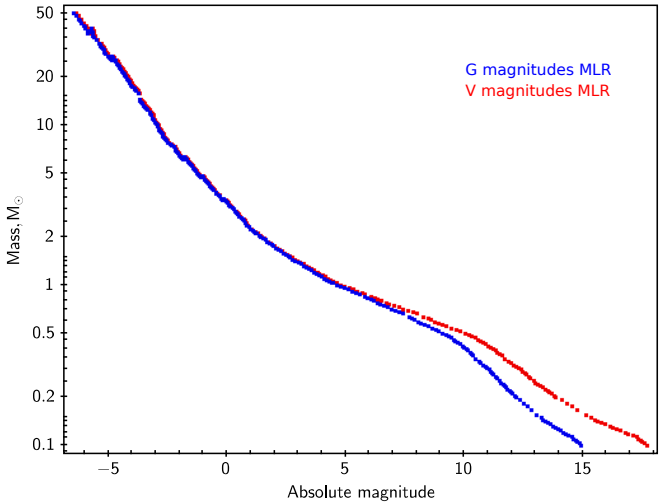}
\caption{Mass-luminosity relations obtained in Section
\ref{sec:mlr}. The blue and red colours stand for absolute $G$ and
$V$ magnitudes respectively. The tabulated form is available in
Table \ref{tab:mlr} of the Appendix.}
    \label{fig:mlr}
\end{figure}

\subsection{Photometric masses}
\label{sec:photometric_mass}

The derived MLR allows us to estimate masses for our binaries. In
principle, the MLR method is only applicable to MS stars with
single-star evolution history; however, most ORB6 binaries meet
these requirements. Despite our sample being comprised of binary
stars, the large distance between components generally allows
their undisturbed evolution. However, any resolved ORB6 component
on its own can be a contact or semidetached binary, therefore, we
should keep in mind that, in certain cases, the MLR method will
yield incorrect mass estimates. Usually the spectral
classification can reveal non-MS stars. In principle, WDS provides
it for most entries, however, these data are rather inhomogeneous,
therefore we choose to calculate photometric masses for all
entries regardless of the availability of spectral classification.

Unlike Kepler's Third Law dynamical estimates, individual stellar
masses are estimated here instead of the combined system's mass.
Depending on identification type, as discussed in Section
\ref{sec:edr3}, \textit{Gaia} magnitudes can be available for one
or two components. When both magnitudes are known, which is
usually a case for wide systems ($\rho \gtrsim 0.5$ arcsec),
masses are calculated for each component separately. If only one
magnitude is available, two extreme cases are considered to define
a range of possible masses: the first hypothesis is that the
contribution of the secondary component is negligible and the mass
of the binary is defined solely by the primary star; the second
hypothesis is that the unresolved source consists of two stars of
equal brightness, and the magnitude of the individual components
is then assumed to be larger by $2.5\log{2} \approx 0.75$ mag in
comparison to the reported value.

Transformation from apparent to absolute magnitudes is required to
enable the MLR method. First of all, saturation correction for
bright stars with $g<8$ mag is applied as prescribed in
\citet{2021A&A...649A...3R}, although its size does not exceed
0.015 mag and generally can be neglected. Then, for a star with
known apparent $g$ magnitude and parallax $\varpi$ (measured in
arcseconds), the absolute magnitude is calculated from:

\begin{equation}
\label{eq:mag}
    G=g+5+5\log{\varpi}-A_G+\sigma_{\text{MLR}}
\end{equation}

The following procedure is used to calculate interstellar
extinction $A_G$. The colour excess $E_{B-V}$ is estimated with
the Stilism 3D model \citep{2019A&A...625A.135L}, the input
distance estimated as $1/\varpi$. Next, the $V$-band extinction is
calculated as $A_v=R_vE_{B-V}$, $R_v=3.1$
\citep{1999PASP..111...63F, 2011ApJ...737..103S}. Finally,
$A_G/A_V=0.84$ conversion \citep{2019ApJ...870..115B} is applied.
The estimated contribution of extinction is generally low, with
$A_G<0.25$ mag for about 96\% of ORB6 entries. Gauss-distributed
errors $\sigma_{\text{MLR}}=0.4$ mag are added to simulate the MLR
uncertainty, with $10^6$ $\varpi$ values sampled to account for
parallax error. After absolute magnitudes are calculated, the MLR
(Fig. \ref{fig:mlr}) is used to derive photometric masses $M_p$.
While we stick to EDR3 photometry in Eq. \ref{eq:mag}, parallaxes
are borrowed from all available sources (Section \ref{sec:edr3},
\ref{sec:third}, \ref{sec:outer}); thus, different mass estimates
are possible for a given ORB6 entry depending on the $\varpi$
value used. Arithmetic weighted means (Eq. \ref{eq:arithmetic})
are adopted when $\varpi_1$ and $\varpi_2$ are independently
known. The 0.159 and 0.841 quantiles define the confidence
intervals $M_p^-$ and $M_p^+$.

\section{Binary-star masses}
\label{sec:comparison} First, we consider 731 resolved systems
with EDR3 magnitudes known for both components, enabling direct
comparison of dynamical $M_{d}$ and photometric $M_{p}$ masses.
Because Kepler's Third Law provides the total system's mass (Eq.
\ref{eq:kepler}), while MLR-derived masses (Fig. \ref{fig:mlr})
are calculated for individual components, we use the sum of the
primary's and secondary's photometric masses
$M_p=M^1_{p}+M^2_{p}$. Inconsistency between dynamical and
photometric masses can be caused by various reasons including
errors in ORB6 orbital elements, inaccurate parallaxes, unresolved
companions, or advanced phase of stellar evolution leading to
incorrect use of the MLR. In general, it may be hard to recognize
which factor or their combination is responsible for the
inconsistency. Moreover, a decent agreement of $M_d$ and $M_p$ can
occur due to lucky error aggregation.

Throughout the paper, we obtained parallaxes from various
catalogues, some of which are in significant disagreement. Formal
error estimates of dynamical masses are usually rather broad and
several $\varpi$ can be accepted. For a given binary we attempt to
choose $\varpi$ which provides the best agreement of $M_d$ and
$M_p$.  We consider the ratio of dynamical and photometric masses,
$r=\frac{M_d}{M_p}$ or $r=\frac{M_p}{M_d}$, $r>1$, and choose
parallax allowing the lowest $r$. In a favorable case when
parallaxes of various origins are similar
($r_{\text{EDR3}}-r<0.05$), we give a preference to the EDR3
parallax due to its higher accuracy. Moreover, when EDR3 and
third-light parallaxes (Section \ref{sec:third}) are comparable,
that with the lower reported error is chosen. We reiterate that
parallax is only one of several parameters required for mass
estimation. In principle, the orbital parameters
$a^{\prime\prime}$ and $P$ can be adjusted instead of $\varpi$.

\begin{figure}
    \centering
    \includegraphics[width=\columnwidth]{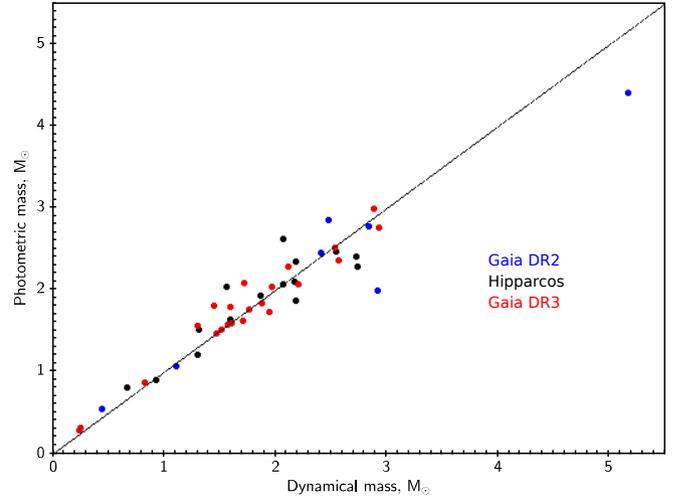}
\caption{Comparison of dynamical and photometric masses calculated
for 44 resolved binaries of grades 1 and 2. The parallax value
that provides the best agreement is selected; its origin is marked
in colour: black stands for original and revised
\textit{Hipparcos} parallaxes; blue -- \textit{Gaia} DR2; red --
for \textit{Gaia} EDR3.}
    \label{fig:grades12}
\end{figure}

We begin with the 44 resolved binaries that have the most reliable
orbits, those of quality grades 1--2, see Fig. \ref{fig:grades12}.
For 10 of these systems, the \textit{Hipparcos} parallaxes are
preferred over those from \textit{Gaia} EDR3. The worst agreement
of masses, $M_d=2.92\pm 0.04 M_\odot$ and $M_p=1.98_{-0.1}^{+0.15}
M_\odot$, is found for WDS 11182+3132 AB ($\xi$ UMa). This star
lacks \textit{Hipparcos} and \textit{Gaia} EDR3 parallaxes,
leaving DR2 as the only available option. Overall, for 10 out of
44 binaries, $M_d$ and $M_p$ do not agree to within the reported
uncertainties.

The rest of the ORB6 binaries have less reliable orbits and hence
larger errors. As discussed in Section \ref{sec:dynamical}, the
error estimates for grades 4--5 are ill-conditioned, therefore the
dynamical mass $M_d^0$, which ignores uncertainties, is introduced
and the ratio $r_0=\frac{M_d^0}{M_p}$, $r_0=\frac{M_p}{M_d^0}$,
$r_0>1$ is additionally calculated. Overall, 326 binaries with
either $r<1.2$ or $r_0<1.2$ are listed in Table \ref{tab:2masses}
of Appendix. While 75\% of orbits with grades 1--3 meet the
threshold, this fraction decreases to 35\% for grades 4--5, see
Table \ref{tab:2mag}. This fact indicates that $M_d$ and $M_p$
disagreements are largely caused by wrong orbital elements.
Solar-type stars are the most common objects among the selected
binaries, as the median primary mass is $M_p^1=1.05 M_\odot$, 90\%
of $M_p^1$ are in the 0.45 -- 2.49 $M_\odot$ range, the
corresponding value for secondary stars is $M_p^2=0.96 M_\odot$ in
the 0.35--1.89 $M_\odot$ range. For 23 binaries
with alternative orbital solutions, we choose those with the
lowest $r$.

Most ORB6 binaries are unresolved in \textit{Gaia} EDR3, and
direct calculation of individual photometric masses is impossible
for them. Instead, two extreme hypotheses are considered assuming
either equal or negligible light contribution of the secondary
relative to the primary component. In the latter case, obviously,
$M_p=M_p^1$. If an unresolved source of the same apparent
magnitude consists of two identical stars, its mass is
approximately 70\% larger, $\dfrac{M_p^=}{M_p^1}\sim 1.7$. Such a
large scatter impedes further analysis, notably the choice of the
best parallax among viable options is hardly possible without
additional assumptions. Any \textit{a priori} information on the magnitude
difference between the components then becomes extremely useful.
For 85 pairs, more than one solution is provided
in ORB6. Since duplicated entries are undesirable, we attempt to
select the best among them with the use of ORB6 $V$ magnitudes. A
complete table containing mass calculations for various parallaxes
is available in Table \ref{tab:master} of the Appendix.

\begin{table}
    \centering
\caption{Agreement of the dynamical $M_d$ and photometric $M_p$
masses  depending on the quality grade of the orbit.
$r=\frac{M_d}{M_p}$ or $r=\frac{M_p}{M_d}$, $r>1$, see Section
\ref{sec:comparison} for details. Resolved systems with known
magnitudes are considered.}
        \label{tab:2mag}

$    \begin{array}{ccccc}
     \text{Grade} &\Sigma& r<1.2 & r<1.1 & r<1.05 \\
                                \hline
1&10&7&7&6\\
2&34&28&20&16\\
3&85&61&44&30\\
4&291&116&85&61\\
5&309&96&64&45\\
\hline
    \end{array}$
\end{table}

\section{Conclusions}
\label{sec:conclusions} In the present paper, we investigate
visual binary stars with known orbits from the ORB6 catalogue and
use them to validate parallax error estimates of \textit{Gaia}
EDR3 and to provide stellar mass estimates for the components of
the pairs. In Section \ref{sec:edr3}, we search EDR3 counterparts
for ORB6 entries and analyze the availability of \textit{Gaia}
astrometric solutions based on the characteristics of the binary
stars (Table \ref{tab:grades}, Fig.
\ref{fig:orb6}-\ref{fig:ruwe}). We show that 2/3 of the EDR3
solutions for systems with projected angular separation
$0.2<\rho<0.5$ arcsec are 2-parameter ones and hence lack parallax
$\varpi$. 521 resolved double stars with $\varpi$ known for both
components are discussed in Section \ref{sec:2plx}. 16 optical
pair candidates are revealed in Section \ref{sec:optical} and
Table \ref{tab:optical}. We show that large discrepancy of EDR3
components' parallaxes reaching $\Delta \varpi /\sigma \sim 18$
does not mean that the system is unbound in Section \ref{sec:
big_errors} and give examples of such binaries in Table
\ref{tab:delta_plx}. A further analysis of reported errors is
described in Section \ref{sec:errors}. We conclude that errors are
underestimated by a factor of 3 for sources with RUWE larger than
2 (Fig. \ref{fig:factor}).

A search for outer companions of ORB6 binaries is made in Section \ref{sec:third} to obtain third-light parallaxes. They are found
to be useful for binaries with 2-parameter solutions, and in some cases they reveal that the \textit{Gaia} parallaxes of the primary and secondary components are unreliable (Table \ref{tab:third_light}).
\textit{Gaia} DR2, TGAS and \textit{Hipparcos} data are added in Section \ref{sec:outer} to supply every ORB6 entry with a parallax. We proceed to dynamical mass estimation in Section
\ref{sec:dynamical}. In Section \ref{sec:mlr}, we derive a synthetic mass-luminosity relation for the $G$ band (Fig. \ref{fig:mlr} and Table \ref{tab:mlr}) which is applied to obtain
photometric masses in Section \ref{sec:photometric_mass}.
Dynamical and photometric masses are calculated for all available
parallaxes of different origin. For resolved binaries, we choose
the $\varpi$ value which provides the best agreement of dynamical
and photometric masses. Overall, for 326 systems, the mass
estimates agree to within 20\% (Table  \ref{tab:2masses}). The
agreement is better for the most reliable orbits (Table
\ref{tab:2mag}, Fig. \ref{fig:grades12}). A complete dataset with
parallaxes and mass estimates for all entries is provided in Table
\ref{tab:master} of the Appendix.

\section*{Acknowledgements}
Authors thank Pavel Kaygorodov, Dana Kovaleva, Rachel Matson, Alexey Rastorguev and Nikolai Samus for the valuable remarks and suggestions. Elaborate review from the referee allowed to improve the paper.
\section*{Data Availability}

The full version of tables is available in the GitHub Repository at \href{https://github.com/chulkovd/ORB6}{https://github.com/chulkovd/ORB6}.

\bibliographystyle{mnras}
\bibliography{example} 

\appendix
\section{Tabulated data}
\label{sec:appendix}

\begin{table}
    \centering
\caption{Mass-luminosity relation for $G$ band derived in Section
\ref{sec:mlr}. Excerpt, full version is available as supplementary material. The plot is shown in
Fig. \ref{fig:mlr}. The approximation formula  $\log M =
0.497 - 0.151G + 0.0106G^2 + 2.48\cdot10^{-4}G^3 -
8.55\cdot10^{-5}G^4 - 4.13\cdot10^{-7}G^5 + 1.93\cdot10^{-7}G^6$
can be used for $-6.4<G<14.9$. The inverse function is $G =
4.98-12.6\log
M+1.84\log^2M+7.22\log^3M-2.31\log^4M-3.49\log^5M+1.59\log^6M$,
$0.1M_\odot<M<50M_\odot$. }
    \begin{tabular}{cc}
Mass, $M_\odot$&$G$\\
    \hline

49.383&-6.399\\
47.535&-6.301\\
45.718&-6.199\\
44.840&-6.183\\
43.144&-6.083\\
        \hline
    \end{tabular}
    \label{tab:mlr}

\end{table}

\begin{table*}
    \centering
\caption{List of resolved ORB6 binaries with decent agreement of
dynamical and photometric masses. Binaries with $r<1.2$ or
$r_0<1.2$ are selected, see Section \ref{sec:comparison} for
details. 326 systems are included. Excerpt, full version is
available as supplementary material. Columns
2--4: parallax (mas) providing the lowest $r$, $r=\frac{M_d}{M_p}$
or $r=\frac{M_p}{M_d}$, $r>1$. Columns 5--7: dynamical mass
($M_\odot$); columns 8--10: total photometric mass; columns
11--13: primary photometric mass; columns 14--16: secondary
photometric mass; columns 17--19: parallax providing the lowest
$r_0$, $r_0=\frac{M_d}{M_p}$ or $r_0=\frac{M_p}{M_d}$, $r_0>1$.
Column 20 is the dynamical mass calculated with uncertainties
neglected. Columns 21--22: angular separation (arcsec) and
position angle ($\degree$).} 
$\begin{array}{cccccccccccccccccccc}
1&2&3&4&5&6&7&8&9&10&11&12&13&14&15&16\\                         \hline
\text{WDS}&\text{Origin of}  \varpi&\varpi&\sigma&M_d^-&M_d&M_d^+&M_p^-&M_p&M_p^+&M_p^{1-}&M_p^1&M_p^{1+}&M_p^{2-}&M_{p}^2&M_p^{2+}\\
00014+3937&\text{H97}&20.42&1.91&1.24&1.70&2.38&1.61&1.72&1.84&0.84&0.90&0.95&0.77&0.83&0.88\\
00021-6817&\text{H97}&63.03&1.98&0.13&1.06&5.8&1.03&1.11&1.19&0.56&0.60&0.65&0.46&0.50&0.54\\
00028+0208&\text{DR3}&23.35&0.02&0.26&2.11&11.7&1.82&1.96&2.11&1.02&1.11&1.21&0.79&0.85&0.90\\
00047+3416&\text{H97}&6.13&1.59&1.01&3.90&18.0&3.44&4.06&5.28&1.78&2.09&2.76&1.67&1.97&2.51\\
00048+3810&\text{DR3}&10.97&0.04&0.64&1.92&6.27&1.77&1.88&1.98&0.92&0.97&1.03&0.85&0.90&0.95\\
\hline
1&17&18&19&20&21&22\\
\hline
\text{WDS}&\text{Origin of}  \varpi&\varpi&\sigma&M_d^0&\rho&\text{PA} \\
00014+3937&\text{H07}&20.15&0.89&1.76&1.33&167\\
00021-6817&\text{DR3}&58.98&0.02&1.19&4.22&131\\
00028+0208&\text{DR3}&23.35&0.02&1.95&1.57&160\\
00047+3416&\text{H07}&5.64&1.42&4.54&0.73&142\\
00048+3810&\text{DR3}&10.97&0.04&1.89&0.89&28\\
        \hline
    \end{array}$
\label{tab:2masses}
\end{table*}

\begin{table*}
    \centering
\caption{Master table containing parallaxes and mass estimates for
all 3460 ORB6 entries. Excerpt, full version is available
as supplementary material. The lines shown are
varied to avoid blank rows. Note that spurious entries with
unrealistic masses are not removed from the table. Columns 1--20
comprise main orbital parameters and primary component
identification in \textit{Gaia} EDR3. Column 1: WDS designation;
column~2: discoverer designation according to ORB6; columns 3--7:
grade, orbital period (in years), angular semimajor axis (mas),
and reported uncertainties according to ORB6. We remind that, for
further calculations, the 75\% quartile (Table \ref{tab:grades}) for the corresponding grade
is applied when error value is absent. Columns 8--12:
\textit{Gaia} EDR3 identification, $G$ apparent magnitude,
parallax with uncertainty (mas), and RUWE of the primary star.}
$\begin{array}{cccccccccccc}
1&2&3&4&5&6&7&8&9&10&11&12\\
\hline
\text{WDS}&\text{Designation}&\text{Grade}&P&\sigma_P&a^{\prime\prime}&\sigma_{a\prime\prime}&\text{EDR3 Identification}&g_1&\varpi_1&\sigma_1&\chi_1\\ \hline
00000-1930&\text{LTT 9831}&9&1.37&0.05&0.0143&0.0028&2341871673090078592&8.94&26.80&0.51&29.4\\
00003-4417&\text{I  1477}&5&384&23&1.023&0.096&4994581292009978112&6.11\\
00006-5306&\text{HJ 5437}&5&904&363&2.80&1.04&4972326695628963584&6.42&16.53&0.02&1.1\\
00008+1659&\text{BAG  18}&5&66.6&&0.531&&2772904695310603520&8.54&33.26&0.04&1.6\\
00014+3937&\text{HLD  60}&3&217&17&0.879&0.018&2881804450094712192&8.95&19.34&0.02&1.1\\
\hline
\end{array}$
\contcaption{Secondary component in EDR3, left blank when pair is
unresolved. Columns 13--17: \textit{Gaia} EDR3 identification, $G$
apparent magnitude, parallax with uncertainty (mas), and RUWE of
the secondary star; columns 18--19: angular separation (arcsec)
and the position angle ($\degree$) according to EDR3 coordinates;
column 20: disagreement of components' parallaxes (Eq
\ref{eq:significance}); column 21: critical mass (Eq.
\ref{eq:escape}, $M_\odot$); columns 22--23: average EDR3 parallax
(Eq. \ref{eq:arithmetic}, mas).} $\begin{array}{cccccccccccc}
1&13&14&15&16&17&18&19&20&21&22&23\\
\hline
\text{WDS}&\text{EDR3 Identification}&g_2&\varpi_2&\sigma_2&\chi_2&\rho&\text{PA}&\Delta\varpi/\sigma&M_e&\varpi_{\rm EDR3}&\sigma_{\rm EDR3} \\
\hline
00000-1930&&&&&&&&&&26.80&0.51\\
00003-4417\\
00006-5306&4972326695627083136&9.57&16.54&0.04&1.6&1.40&337&0.18&1.48&16.53&0.02\\
00008+1659&&&&&&&&&&33.26&0.04\\
00014+3937&2881804450094712320&9.43&20.03&0.04&1.8&1.33&167&16.35&0.17&19.50&0.02\\
\hline
\end{array}$
\contcaption{Third-light component in EDR3 (see Section
\ref{sec:third} for details). Columns 24--28: \textit{Gaia} EDR3
identification, $G$ apparent magnitude, parallax with uncertainty
(mas), and RUWE; column 29: angular distance (arcsec) from the
primary component; column 30: projected linear separation
$\rho_3/\varpi_3$ ($10^3$ AU); column 31:
disagreement of primary and third-light parallax, Eq
$\ref{eq:significance}$; column 32: critical mass (Eq.
\ref{eq:escape}, $M_\odot$).}
$\begin{array}{ccccccccccc}
1&24&25&26&27&28&29&30&31&32\\
\hline
\text{WDS}&\text{EDR3 Identification}&g_3&\varpi_3&\sigma_3&\chi_3&\rho_3&a_3&\Delta\varpi/\sigma&M_e&\\
\hline
00003-4417&4994581498167873152&17.68&12.78&0.11&1.0&40.4&3.16&\\
00024+1047&2765432654808342016&8.41&10.08&0.15&4.9&63.4&6.29\\
00046+4206&384361163100177280&9.96&4.43&0.02&1.2&5.3&1.21&4.03&5.00\\
00047+3416&2875176250406193920&10.44&4.70&0.05&2.8&95.9&20.4&1.47&10.6\\
00057+4549&386653747925624576&8.30&86.82&0.03&1.2&6.0&0.069&2.03&0.23\\
\hline
\end{array}$
\label{tab:master}
\end{table*}

\begin{table*}
    \centering

\contcaption{Identification in \textit{Gaia} DR2. Columns 33--38:
identification, parallax with corresponding uncertainty for
primary and secondary component (mas). Columns 39--40: average DR2
parallax (Eq. \ref{eq:arithmetic}, mas). Sources without
available parallax are not provided.  }
$\begin{array}{ccccccccccc}
1&33&34&35&36&37&38&39&40\\
\hline
\text{WDS}&\text{Primary identification}&\varpi_1&\sigma_1&\text{Secondary identification}&\varpi_2&\sigma_2&\varpi_{\rm DR2}&\sigma_{\rm DR2}\\
\hline
00000-1930&2341871673090078592&25.12&0.32&&&&25.12&0.32\\
00003-4417&4994581292009978112&8.14&0.66&&&&8.14&0.66\\
00006-5306&4972326695628963584&16.35&0.04&4972326695627083136&16.63&0.25&16.36&0.04\\
00008+1659&2772904691015625984&33.16&0.11&&&&33.16&0.11\\
00014+3937&2881804450094712192&19.27&0.07&2881804450094712320&19.68&0.12&19.37&0.06\\
\hline
\end{array}$
\contcaption{Identification in TGAS. Columns 41--46: identification,
parallax with corresponding uncertainty for primary and secondary component (mas). Columns 47--48: average
TGAS parallax (Eq. \ref{eq:arithmetic}, mas). }
$\begin{array}{ccccccccccc}
1&41&42&43&44&45&46&47&48\\
\hline
\text{WDS}&\text{id}_1&\varpi_1&\sigma_1&\text{id}_2&\varpi_2&\sigma_2&\varpi_{\rm TGAS}&\sigma_{\rm TGAS}\\
\hline
00494-2313&2348830512245670912&50.37&0.46&&&&50.37&0.46\\
00507+6415&524013669703057536&5.90&0.96&&&&5.90&0.96\\
00524-6930&4691995687749952384&16.53&0.25&4691995996987597568&14.89&0.35&15.98&0.20\\
00542+4318&375705975069410176&5.25&0.24&&&&5.25&0.24\\
00569-5153&4928347428812956416&24.54&0.52&&&&24.54&0.52\\
\hline
\end{array}$
\contcaption{Identification in \textit{Hipparcos}. Columns 49--54:
identification of primary component, Johnson $V$ apparent
magnitude, original \citep{1997A&A...323L..49P} and revised
\citep{2007ASSL..350.....V} parallax with uncertainty (mas).
Columns 55--60: the corresponding values for secondary component;
columns 61--64: average original and revised \textit{Hipparcos}
parallax (Eq. \ref{eq:arithmetic}, mas).  }
$\begin{array}{cccccccccccccccccc}
1&49&50&51&52&53&54&55&56&57&58&59&60&61&62&63&64\\
\hline
\text{WDS}&\text{Hip}_1&\text{mag}_1&\varpi_1^{97}&\sigma_1^{97}&\varpi_1^{07}&\sigma_1^{07}&\text{Hip}_2&\text{mag}_2&\varpi_2^{97}&\sigma_2^{97}&\varpi_2^{07}&\sigma_2^{07}&\varpi_{\rm H97}&\sigma_{\rm H97}&\varpi_{\rm H07}&\sigma_{\rm H07}\\
\hline
00053-0542&443&4.61&25.38&1.05&25.32&0.53&&&&&&&25.38&1.05&25.32&0.53\\
00055+3406&461&7.86&11.04&0.91&10.30&0.75&&&&&&&11.04&0.91&10.3&0.75\\
00057+4549&473&8.20&85.10&2.74&88.44&1.56&428&9.95&86.98&1.41&88.88&1.42&86.59&1.25&88.68&1.05\\
00059+1805&495&8.58&25.77&2.07&26.92&1.20&&&&&&&25.77&2.07&26.92&1.20\\
00061+0943&510&7.80&11.07&1.00&11.52&0.93&&&&&&&11.07&1.00&11.52&0.93\\
\hline
\end{array}$

\contcaption{Extraneous parallaxes applicable for pairs with
absent \textit{Gaia} and \textit{Hipparcos} data and auxiliary
information. Columns 65--67: parallax, its uncertainty,
and reference (usually linked to VizieR). When $\sigma_x$ is not
reported, $\sigma_x/\varpi_x=0.2$ is adopted for further
calculations. Columns 68--72 are Boolean markers; column 68:
identical WDS designation exists, applicable both to multiple
systems and multiple solutions for the same pair; column 69:
duplicated orbits for the same pair; column 70: best entry choice
applicable to remove duplicated orbits; column 71: optical pair
candidate (Section \ref{sec:optical}); column 72: merged source
indicator applicable when ORB6 component is obscured by a bright
star in a close vicinity. Photometric mass for such systems is
unreliable.} $\begin{array}{cccccccccccccccccc}
1&65&66&67&68&69&70&71&72\\
\hline
\text{WDS}&\varpi_x&\sigma_x&\text{Reference}&b_i&b_d&b_b&b_0&b_m\\
\hline
00114+5850&1.67&0.33&\text{J/ApJ/653/657}&1&1&0&0&0\\
00114+5850&1.67&0.33&\text{J/ApJ/653/657}&1&1&0&0&0\\
00114+5850&1.67&0.33&\text{J/ApJ/653/657}&1&1&1&0&0\\
00114+5850&1.67&0.33&\text{J/ApJ/653/657}&1&1&0&0&0\\
00152+2722&&&&0&0&0&1&0\\
00431+7659 & 7.74 & 1.55 & \text{J/AJ/156/102}&0&0&1&0&0\\
06298-5014 & 19.44 & 0.66 & \text{Revised \textit{Hipparcos} parallax of AB pair}&1&0&1&0&0  \\
11182+3132 & 114.49&0.43& \text{\textit{Gaia} DR2 parallax for B component}&1&0&1&0&0\\
22385-1519&293.6&0.9&\text{2010A\&ARv..18...67T}&1&0&1&0&1\\
22385-1519&293.6&0.9&\text{2010A\&ARv..18...67T}&1&0&1&0&0\\
\hline
    \end{array}$
\end{table*}

\begin{table*}
    \centering

\contcaption{Dynamical and photometric masses (measured in
$M_\odot$) calculated with \textit{Gaia} EDR3 parallaxes. $M_0^d$:
dynamical mass  errors neglected. $M_d$: median dynamical mass
with confidence interval $M_d^-$ and $M_d^+$ (Section
\ref{sec:dynamical}). $M_p^1$: median photometric mass (Section
\ref{sec:photometric_mass}) of primary component with confidence
interval $M_p^{1-}$ and $M_p^{1+}$. $M_p^{2-}$, $M_p^2$ and
$M_p^{2+}$ are the corresponding values for secondary component.
$M_p^2$ is absent for unresolved pairs. $M_p^=$ with error
estimate $M_p^{=-}$ and $M_p^{=+}$ is the total mass of unresolved
binary in assumption of equal contribution from components. Total
mass $M_p$ of unresolved binary is therefore confined between
$M_p^1$ and $M_p^=$ depending on relative magnitude contrast of
components. $M_p^-$,$M_p$,$M_p^+$ is total photometric mass of
resolved binaries. } $\begin{array}{ccccccccccccccccccc}
1&73&74&75&76&77&78&79&80&81&82&83&84&85&86&87&88\\
\hline
\text{WDS}&M_d^0&M_d^-&M_d&M_d^+&M_p^{1-}&M_p^1&M_p^{1+}&M_p^{2-}&M_p^2&M_p^{2+}&M_p^{=-}&M_p^=&M_p^{=+}&M_p^-&M_p&M_p^+ \\
\hline
  00000-1930&0.08&0.04&0.08&0.14&0.76&0.81&0.86&&&&1.36&1.44&1.53\\
  00003-4417\\
  00006-5306&5.97&1.34&6.17&24.0&1.41&1.54&1.69&0.81&0.86&0.92&&&&2.22&2.41&2.61\\
  00008+1659&0.92&0.12&1.00&5.48&0.75&0.80&0.86&&&&1.35&1.43&1.51\\
  00014+3937&1.94&1.66&1.94&2.30&0.86&0.91&0.96&0.79&0.84&0.89&&&&1.64&1.75&1.85\\
\hline
\end{array}$
\contcaption{Dynamical and photometric masses calculated with
third-light EDR3 parallaxes. }

$\begin{array}{ccccccccccccccccccc}
1&89&90&91&92&93&94&95&96&97&98&99&100&101&102&103&104\\
\hline
\text{WDS}&M_d^0&M_d^-&M_d&M_d^+&M_p^{1-}&M_p^1&M_p^{1+}&M_p^{2-}&M_p^2&M_p^{2+}&M_p^{=-}&M_p^=&M_p^{=+}&M_p^-&M_p&M_p^+ \\
\hline
  00003-4417&3.48&2.53&3.48&4.68&1.72&1.90&2.10&&&&2.90&3.17&3.48\\
  00024+1047&2.85&0.97&2.90&9.49&1.12&1.22&1.32&&&&1.96&2.08&2.26\\
  00046+4206&10.5&8.51&10.5&13.0&3.68&4.29&4.88&&&&5.76&6.69&7.51\\
  00047+3416&7.87&2.67&8.01&26.2&2.21&2.59&2.99&2.07&2.29&2.71&&&&4.28&4.88&5.70\\
  00057+4549&1.91&0.25&2.07&11.4&0.56&0.60&0.65&0.49&0.53&0.56&&&&1.06&1.13&1.21\\
    \hline
\end{array}$
\contcaption{Dynamical and photometric masses calculated with \textit{Gaia} DR2 parallaxes.}
$\begin{array}{ccccccccccccccccccc}
1&105&106&107&108&109&110&111&112&113&114&115&116&117&118&119&120\\
\hline
\text{WDS}&M_d^0&M_d^-&M_d&M_d^+&M_p^{1-}&M_p^1&M_p^{1+}&M_p^{2-}&M_p^2&M_p^{2+}&M_p^{=-}&M_p^=&M_p^{=+}&M_p^-&M_p&M_p^+ \\
\hline
  00000-1930&0.10&0.05&0.10&0.17&0.78&0.83&0.88&&&&1.39&1.47&1.57\\
  00003-4417&13.4&9.07&13.5&19.9&2.20&2.61&3.05&&&&3.61&4.06&4.55\\
  00006-5306&6.16&1.39&6.37&24.8&1.42&1.55&1.70&0.82&0.87&0.92&&&&2.24&2.42&2.62\\
  00008+1659&0.93&0.12&1.00&5.54&0.75&0.80&0.86&&&&1.35&1.43&1.52\\
  00014+3937&1.98&1.69&1.98&2.34&0.86&0.91&0.96&0.79&0.84&0.89&&&&1.65&1.75&1.85\\
\hline
\end{array}$
\contcaption{Dynamical and photometric masses calculated with \textit{Gaia} TGAS parallaxes.}
$\begin{array}{ccccccccccccccccccc}
1&121&122&123&124&125&126&127&128&129&130&131&132&133&134&135&136\\
\hline
\text{WDS}&M_d^0&M_d^-&M_d&M_d^+&M_p^{1-}&M_p^1&M_p^{1+}&M_p^{2-}&M_p^2&M_p^{2+}&M_p^{=-}&M_p^=&M_p^{=+}&M_p^-&M_p&M_p^+ \\
\hline
00494-2313&1.00&0.73&1.01&1.34&0.84&0.89&0.94&&&&1.48&1.58&1.70\\
00507+6415&3.47&2.12&3.48&6.12&3.91&4.73&5.83&&&&6.08&7.14&8.85\\
00524-6930&3.58&1.56&3.50&12.2&1.39&1.52&1.67&1.21&1.32&1.43&&&&2.61&2.84&3.10\\
00542+4318&2.70&0.35&2.92&16.2&1.58&1.74&1.93&&&&2.68&2.93&3.21\\
00569-5153&2.01&1.68&2.01&2.40&0.80&0.86&0.91&&&&1.43&1.52&1.62\\
\hline
\end{array}$
\contcaption{Dynamical and photometric masses calculated with original \textit{Hipparcos} \citep{1997A&A...323L..49P} parallaxes.  }
$\begin{array}{ccccccccccccccccccc}
1&137&138&139&140&141&142&143&144&145&146&147&148&149&150&151&152\\
\hline
\text{WDS}&M_d^0&M_d^-&M_d&M_d^+&M_p^{1-}&M_p^1&M_p^{1+}&M_p^{2-}&M_p^2&M_p^{2+}&M_p^{=-}&M_p^=&M_p^{=+}&M_p^-&M_p&M_p^+ \\
\hline
  00053-0542&0.15&0.06&0.15&0.29&1.84&2.06&2.29&&&&3.08&3.38&3.75\\
  00055+3406&0.88&0.65&0.88&1.22&1.28&1.40&1.55&&&&2.20&2.41&2.63\\
  00057+4549&1.1&0.37&0.37&1.12&0.56&0.61&0.65&0.56&0.60&0.65&&&&1.13&1.21&1.30\\
  00057+4549&1.50&0.88&1.51&2.62&0.56&0.61&0.65&0.56&0.60&0.65&&&&1.13&1.21&1.30\\
  00057+4549&1.93&0.25&2.08&11.5&0.56&0.60&0.65&0.50&0.53&0.56&&&&1.06&1.13&1.21\\
  \hline
\end{array}$
\contcaption{Dynamical and photometric masses calculated with revised \textit{Hipparcos} \citep{2007ASSL..350.....V} parallaxes.   }
$\begin{array}{ccccccccccccccccccc}
1&153&154&155&156&157&158&159&160&161&162&163&164&165&166&167&168\\
\hline
\text{WDS}&M_d^0&M_d^-&M_d&M_d^+&M_p^{1-}&M_p^1&M_p^{1+}&M_p^{2-}&M_p^2&M_p^{2+}&M_p^{=-}&M_p^=&M_p^{=+}&M_p^-&M_p&M_p^+ \\
\hline
  00053-0542&0.15&0.06&0.15&0.29&1.85&2.06&2.29&&&&3.09&3.39&3.75\\
  00055+3406&1.09&0.81&1.09&1.46&1.33&1.45&1.60&&&&2.27&2.48&2.71\\
  00057+4549&0.98&0.33&1.00&3.26&0.56&0.60&0.64&0.55&0.59&0.64&&&&1.11&1.19&1.28\\
  00057+4549&1.33&0.79&1.34&2.32&0.56&0.60&0.64&0.55&0.59&0.64&&&&1.11&1.19&1.28\\
  00057+4549&1.79&0.24&1.94&10.7&0.56&0.60&0.64&0.49&0.52&0.55&&&&1.05&1.12&1.20\\
  \hline
\end{array}$
\contcaption{Dynamical and photometric masses calculated with extraneous parallaxes.} $\begin{array}{ccccccccccccccccccc}
1&169&170&171&172&173&174&175&176&177&178&179&180&181&182&183&184\\
\hline
\text{WDS}&M_d^0&M_d^-&M_d&M_d^+&M_p^{1-}&M_p^1&M_p^{1+}&M_p^{2-}&M_p^2&M_p^{2+}&M_p^{=-}&M_p^=&M_p^{=+}&M_p^-&M_p&M_p^+ \\
\hline 
  00114+5850&2.77&0.84&2.97&11.9&0.72&0.79&0.87&&&&1.30&1.41&1.55\\
  00431+7659&10.2&1.29&11.5&74.9&1.05&1.18&1.35&0.98&1.08&1.23&&&&2.03&2.26&2.57\\
  00520+3154&0.92&0.52&0.93&1.87&0.91&0.98&1.09&0.93&1.01&1.13&&&&1.83&1.99&2.22\\
  06298-5014&2.05&1.83&2.06&2.31&0.87&0.93&0.98&&&&1.55&1.66&1.76\\
  22385-1519&0.33&0.32&0.33&0.33&0.14&0.15&0.17&&&&0.23&0.25&0.28\\
  \hline
  \end{array}$
  \end{table*}

\bsp    
\label{lastpage}
\end{document}